\documentclass[a4paper,11pt
]{article}
\pdfoutput=1

\usepackage{cpcpub}

\usepackage[T1]{fontenc}

\usepackage{slashed}

\usepackage{threeparttable}

\usepackage{caption}

\usepackage{subfigure}

\title{\boldmath Observational constraints on dark matter decaying via gravity portals}

\author{Sun Xu--Dong}
\author[1]{and Dai Ben--Zhong\note{Corresponding author.}}

\affiliation{School of Physics and Astronomy, Yunnan University, Kunming, 650091, China}
\affiliation{Key Laboratory of Astroparticle Physics, Yunnan Province, Kunming 650091, China.}

\emailAdd{bestsunxudong@126.com}
\emailAdd{bzhdai@ynu.edu.cn}

\abstract{
Global symmetry can guarantee the stability of dark matter particles (DMps). However, the nonminimal coupling between dark matter (DM) and gravity can destroy the global symmetry of DMps, which in turn leads to their decay. Under the framework of nonminimal coupling between scalar singlet dark matter (ssDM) and gravity, it is worth exploring to what extent the symmetry of ssDM is broken. It is suggested that the total amount of decay products of ssDM cannot exceed current observational constraints. Along these lines, the data obtained with satellites such as Fermi-LAT and AMS-02 can limit the strength of the global symmetry breaking of ssDM. Since the mass of many well--motivated DM candidates may be in the GeV--TeV range, we determine a reasonable parameter range for the lifetime in this range. We find that when the mass of the ssDM is around the electroweak scale (246 GeV), the corresponding 3--$\sigma$ lower limits of the lifetime of ssDM is $5.3\times10^{26}$ s. Our analysis of ssDM around the typical electroweak scale contains the most abundant decay channels of all mass range, so the analysis of the behaviour of ssDM under the influence of gravity is more comprehensive.
}


\arxivnumber{2002.09955}

\begin{document}
\maketitle
\flushbottom

\section{Introduction}

Observations of the rotation curves of galaxies, the Bullet Cluster, gravitationally lensed galaxy clusters, type Ia supernovae, baryonic acoustic oscillations, and anisotropies in the cosmic microwave background have all implied the existence of dark matter (DM)~\cite{PDM}. The Standard Model of particle physics describes electromagnetism, as well as the weak and strong nuclear forces successfully~\cite{reviewPP}, however it does not currently accommodate the existence of any dark matter particles (DMps). All these imply that physics beyond the Standard Model should be in place~\cite{yu1}~\cite{yu2}.

Among the various properties of DMps, we are concerned with their stability, because if DMps are unstable, we could observe their decay products with satellites~\cite{stabilityofDM}~\cite{yu3}~\cite{yu5}. The stability of electrons is guaranteed by electric charge conservation, while the stability of neutrinos is guaranteed by Lorentz symmetry. Similarly, current observations have suggested that DM is stable and may be composed of particles. It is usually assumed that DMps have global symmetry in Minkowski space-time, such as the hypothetical $Z_2$ symmetry~\cite{singletscalars}~\cite{ScalarPhantoms}. But every particle is subject to gravitational interactions. In reality, there is no Minkowski space-time, and gravity does not necessarily couple to DM minimally. In the minimal coupling regime, matter distribution decides the distribution of gravitons, and gravitons and matter do not transform each other. However, if gravitons couples to DM nonminimally, the global symmetry of DMps can be broken~\cite{SandSinFTandG}~\cite{GandGSym}. Consequently, the stability of DMps is no longer preserved under the influence of gravity~\cite{violationRparity}~\cite{PlanckscaleGS}~\cite{PlanckscaleWIMP}~\cite{DMandGS}, implying that DMps could decay via nonminimal coupling to gravity.

O. Cat\`{a} et al.~\cite{Oscar}~\cite{DMdecayTGP} give such models of scalar singlet dark matter (ssDM), inert doublet DM and fermionic DM with global symmetry breaking induced by nonminimal coupling to gravity. There were also other attempts to study the nonminimal coupling regime. For example, the Higgs field may have nonminimal coupling with gravity in Higgs inflation~\cite{Higgsinflation}. If the mass of the dark matter particle (DMp) is less than 270~MeV, such a particle could concurrently be acting as an inflaton~\cite{inflatonasDM}. There are also nonminimal coupling models of DM and gravity where the global symmetry is not destroyed by gravity~\cite{ProbingGDM}~\cite{scalarDMfromIF}~\cite{ScaleInvariantScalar}.
As both an inflaton and DMp, the nonminimal coupling between a complex scalar field and gravity has also been used to explain the electroweak phase transition~\cite{inflationEWandDM}~\cite{NscalarinflationEWandDM}.

Many observations and experiments could set constraints on the strength of the global symmetry breaking of DMps. Currently, there are many types of experiments and observational methods being used to search for DMps. Direct detection methods rely on monitoring nucleon recoil induced by interactions with DMps distributed around the Earth~\cite{DirectWIMPs}. Indirect detection methods search for photons, neutrinos, and/or cosmic rays produced by DMps using satellites and Earth-based instrumentation~\cite{IndirectSearch}. The Large Hadron Collider also serves as a complementary experiment in the search for DM. Cosmological studies have also provided constraints on DM. If nonminimal coupling to gravity breaks the global symmetry of DMps, DM would be unstable, and it would consequently decay into observable particles such as cosmic rays~\cite{AMS02atlast}, neutrinos~\cite{Sun79IceCube} or cosmic gamma-rays~\cite{SixYearsFermiLAT}. So while no conclusive particle signal has yet been attributed to DM~\cite{newera}, current observations can still be used to set constraints on the stability of DMp.

Using chiral perturbation theory, O. Cat\`{a} et al.~\cite{Sharpfeature} provided the allowed parameter space of light ssDM particles less massive than 1~GeV, in which the decay products have a sharp photon spectrum. These authors obtained the strongest constraints to date using Fermi-LAT gamma-ray observations. However, the mass range of weakly interacting massive particles (WIMP) and super WIMPs proposed based on the gauge hierarchy problem as well as hidden DM based on the gauge hierarchy problem and new flavour physics is expected to be GeV--TeV~\cite{DMcandidates}. And, if the mass of the DMp is in the GeV--TeV range, more decay channels will be opened and the decay properties of DMps will be quite diverse. Assuming that the lifetime of DMps is longer than the age of the universe and using observation data from neutrino telescopes, O. Cat\`{a} et al.~\cite{Oscar}~\cite{DMdecayTGP} provided rough restrictions of the nonminimal coupling coefficient between the ssDM, the inert doublet DM, the fermionic DM and the Ricci scalar around the GeV--TeV range.

In the case of DM decay, constraints obtained via indirect-detection methods play an important role. As indirect-detection methods, satellites such as Fermi-LAT~\cite{FermiLAT}, Alpha Magnetic Spectrometer (AMS)~\cite{AMS02}, and DArk Matter Particle Explorer (DAMPE)~\cite{DAMPE} can obtain sensitive observations of high-energy photons and cosmic rays. Given that DAMPE is unable to distinguish positrons from electrons, in this current work we only consider positron data obtained by AMS-02~\cite{AMS02} and photon data obtained by Fermi-LAT~\cite{FermiLAT} to yield conservative indirect restrictions of the GeV--TeV range.

According to the work of O. Cat\`{a} et al.~\cite{DMdecayTGP}, the action is constructed in the Jordan frame. When using Feynman diagrams to calculate the specific decay channel, one can choose to calculate it in either Jordan frame or Einstein frame. For example, J. Ren et al.~\cite{JFandEF} used the quantum field theory method to calculate Higgs inflation both in Jordan frame and Einstein frame. They obtained the same result using both, which reflects the equivalence of the Jordan frame and the Einstein frame in these scenarios. Then, in the Einstein frame, we calculate the spectra of photons and positrons arising from the decay of ssDM particles in the GeV--TeV range where WIMPs, super WIMPs and hidden DM mass may likely be. Finally, we obtain constraints on the lifetime and the nonminimal coupling constant $\xi$, which reflects the strength of the global symmetry breaking of ssDM particles, by comparing our theoretical spectra to observations made by Fermi-LAT and AMS-02.

The structure of this paper is as follows. In Section~\ref{MandBR}, we introduce the model and discussed the decay branch ratio of ssDM around the electroweak scale. In Section~\ref{decayspectrumGSB}, we describe the calculation of the ssDM decay spectrum induced by global symmetry breaking. In Section~\ref{ConstraintsfromAMSandFermiLAT}, we show the statistic methods to compare the expected spectrum from decaying ssDM with the observed spectrum from Fermi-LAT and AMS-02. In Section~\ref{Results}, we give decay spectra of ssDM induced by global symmetry breaking and the reasonable parameter space of the lifetime and the nonminimal coupling constant. The discussion and conclusions are presented in Section~\ref{DandC}.

\section{The Model and Branch Ratio}\label{MandBR}

\subsection{The Model}

O. Cat\`{a} et al.~\cite{Oscar} considered that DM can couple to the Ricci scalar nonminimally and whose global symmetry is broken in curved space-time. In this paper, we focus on ssDM. In Jordan Frame, the action $\mathcal{S}$ of system can be written as:

\begin{equation}
\mathcal{S}
=\int d^4x \sqrt{-g} [-\frac{R}{2\kappa^2}+\mathcal{L}_{SM}+\mathcal{L}_{DM} -\xi M \varphi R ]
\label{scalarsinglet}
\end{equation}
where $g$ is the determinant of metric tensor $g_{\mu\nu}$.

The Einstein--Hilbert Lagrangian $-R/2\kappa^2$ describes the gravitational sector, where $R$ is the Ricci scalar and $\kappa=\sqrt{8\pi G}$ is the inverse (reduced) Planck mass with $G$ the Newtonian gravitational constant.

$\mathcal{L}_{SM}$ is the Standard Model Lagrangian. It accurately describes the electromagnetism, weak and strong nuclear forces at energies around the electroweak scale and could be cast as,
\begin{equation}
\mathcal{L}_{SM}=\mathcal{T}_F+\mathcal{T}_f+\mathcal{T}_H+\mathcal{L}_Y-\mathcal{V}_H
\end{equation}
where $\mathcal{V}_H$ is the Higgs potential, $\mathcal{L}_Y$ is the Yukawa interaction term and $\mathcal{T}_i$ are the kinetic terms of spin--one particles, fermions and scalars,
\begin{subequations}
\begin{equation}
\mathcal{T}_F=-\frac{1}{4}g^{\mu\nu}g^{\lambda\rho}F^a_{\mu\lambda}F^a_{\nu\rho}
\end{equation}
\begin{equation}
\mathcal{T}_{f}=\frac{i}{2} \bar{f} \stackrel{\leftrightarrow}{\slashed{\nabla}} f
\end{equation}
\begin{equation}
\mathcal{T}_H=g^{\mu\nu}(D_\mu \phi)^\dagger (D_\nu \phi).
\end{equation}
\end{subequations}
In these equations, the slashed derivative operator is defined as $\slashed{\nabla} = \gamma^a e^\mu_a \nabla_\mu$, where $\nabla_\mu = D_\mu - \frac{i}{4} e^b_\nu (\partial_\mu e^{\nu c})\sigma_{bc}$ and $e^{\nu c}$ is the vierbein. $D_\mu$ represents the gauge covariant derivative. $\phi$ denotes the Higgs doublet.

In Eq.~\eqref{scalarsinglet}, $\mathcal{L}_{DM}=\mathcal{T}_\varphi-V(\varphi,X)$ is the Lagrangian of the ssDM, where $\varphi$ represents ssDM. $V(\varphi,X)$ is the DM potential. Since the DM potential contains interatctions between ssDM and Standard Model particles $X$, it could be responsible for the correct DM relic abundance.

The research content of this paper comes from the last term of Eq.~\eqref{scalarsinglet}. Specifically, $-\xi M \varphi R $ is the assumed non--minimal coupling operator between the ssDM and gravity, where $\xi$ is the coupling constant, $M$ is a parameter with dimension one so that $\xi$ is dimensionless. For convenience, we fix $M=\kappa^{-1}$. This non--minimal coupling operator breaks global $\mathbb{Z}_2$ symmetry of $\varphi$, consequently induces ssDM decay into Standard Model particles.

Using conformal transformation, as shown in Eq.~\eqref{conformaltransformation}:
\begin{equation}
\tilde{g}_{\mu\nu} = \Omega^2 g_{\mu\nu}
\label{conformaltransformation}
\end{equation}
where $\Omega^2=1+2\xi M\kappa^2 \varphi$, one can acquire action in Einstein Frame, which is shown as Eq.~\eqref{actionEinsteinFrame}:
\begin{equation}
\mathcal{S}
=\int d^4x \sqrt{-\tilde{g}} [
-\frac{\tilde{R}}{2\kappa^2}
+\frac{3}{\kappa^2} \frac{\Omega_{,\rho}\tilde{\Omega}^{,\rho}}{\Omega^2}
+\tilde{\mathcal{L}}_{SM}+\tilde{\mathcal{L}}_{DM}]
\label{actionEinsteinFrame}
\end{equation}
where:
\begin{equation}
\tilde{\mathcal{L}}_{SM}=\tilde{\mathcal{T}}_F+\Omega^{-3}\tilde{\mathcal{T}}_f+\Omega^{-2} \tilde{\mathcal{T}}_H+\Omega^{-4}(\mathcal{L}_Y-\mathcal{V}_H)
\label{decayoranniterm}
\end{equation}
and $\tilde{\mathcal{L}}_{DM}=\tilde{\mathcal{T}}_\varphi/\Omega^2-V(\varphi,X)/\Omega^{4}$. In these expressions, all tilded quantities are formed from $\tilde{g}_{\mu\nu}$.

Eq.~\eqref{decayoranniterm} indicates that DM $\varphi$ could decay or annihilate into Standard Model particles through gravity portals. Taylor expansion of Eq.~\eqref{decayoranniterm} with respect to $\xi$ shows that the dominant term is the decay term, as shown in Eq.~\eqref{decayterm}:
\begin{equation}
\tilde{\mathcal{L}}_{SM,\varphi}
=-2\kappa\xi \varphi
[\frac{3}{2}\tilde{\mathcal{T}}_f
+ \tilde{\mathcal{T}}_H
+2(\mathcal{L}_Y-\mathcal{V}_H)]
\label{decayterm}
\end{equation}
Using Eq.~\eqref{decayterm}, O. Cat\`{a} et al.~\cite{DMdecayTGP} gave Feynman rules for DM decay, as shown in Table~\ref{vertexrules}.
\begin{table}
\centering
\caption{\label{vertexrules} Feynman rules for DM decay}
\begin{tabular}{|l |l|l|}
\hline
terms from $\tilde{\mathcal{L}}_{sm,\varphi}$ \eqref{decayterm}
& physical process
& Feynman rules \\
\hline
$\xi \kappa m_{f_i} \varphi \bar{f}_i f_i $
& $\varphi \rightarrow \bar{f}_i , f_i$
& $i\xi \kappa m_{f_i} $ \\
\hline
$- 3 \xi \kappa \varphi Y_\mu\bar{f}_i (\gamma^a e^\mu_a) ( a_{f_{ij}}-b_{f_{ij}}\gamma^5) f_j $
& $ \varphi \rightarrow Y_\mu, \bar{f}_i, f_j $
& $- 3 i\xi \kappa (\gamma^a e^\mu_a) ( a_{f_{ij}}-b_{f_{ij}}\gamma^5)$ \\
\hline
$- \xi \kappa \varphi [ (\partial_\mu h)^2 - 2 m_h^2 h^2] $
& $ \varphi \rightarrow h,h $
& $ 2i\xi \kappa [p_{1\mu} p_2^\mu + 2 m_h^2 ] $ \\
\hline
$- \xi \kappa \varphi [2m_W^2 W^{\mu +} W_\mu^- + m_Z^2 Z^\mu Z_\mu ] $
& $ \varphi \rightarrow Y_\mu , Y_\nu $
& $-2 i\xi \kappa m_{Y_\mu}^2 \tilde{g}^{\mu\nu} $ \\
\hline
$- 2 \xi \kappa \varphi \frac{h}{v} [2m_W^2 W^{\mu +} W_\mu^- + m_Z^2 Z^\mu Z_\mu ] $
& $ \varphi \rightarrow h, Y_\mu , Y_\nu $
& $- 4 i\xi \kappa \frac{1}{v} m_{Y_\mu}^2 \tilde{g}^{\mu\nu} $ \\
\hline
$- \xi \kappa \varphi \frac{h^2}{v^2} [2m_W^2 W^{\mu +} W_\mu^- + m_Z^2 Z^\mu Z_\mu ]$
& $ \varphi \rightarrow h,h,Y_\mu , Y_\nu $
& $- 4i\xi \kappa \frac{1}{v^2} m_{Y_\mu}^2 \tilde{g}^{\mu\nu} $ \\
\hline
$ 4 \xi \kappa \varphi m_{f_i} \bar{f}_i f_i \frac{h}{v}$
& $ \varphi \rightarrow h,\bar{f}_i , f_i $
& $ 4 i\xi \kappa \frac{m_{f_i}}{v}$ \\
\hline
$2 \xi \kappa \frac{m_h^2}{v} \varphi h^3$
& $ \varphi \rightarrow h,h,h $
& $12 i\xi \kappa \frac{m_h^2}{v}$ \\
\hline
$\frac{1}{2} \xi \kappa \frac{m_h^2}{v^2} \varphi h^4 $
& $ \varphi \rightarrow h,h,h,h $
& $12i \xi \kappa \frac{m_h^2}{v^2} $ \\
\hline
\end{tabular}
\begin{tablenotes}
\item In the table, $f_i$ represents a fermion and index $i$ includes all fermion flavours,
$Y_\mu$ represents a spin-one particle,
$a_{f_{ij}}$ and $b_{f_{ij}}$ can be obtained from the expansion of $\tilde{\mathcal{T}}_f$.
$W^\mu$ represents the W boson and $Z^\mu$ represents the Z boson,
$h$ represents the Higgs boson,
$v=246.2$~GeV is the Higgs vacuum expectation value,
$m_{Y_\mu}$ represents the mass of the spin-one particle,
$m_{f_i}$ represents the mass of the fermion,
$m_h$ represents the mass of the Higgs boson.
The second column lists the decay channels. For example, $\varphi \rightarrow \bar{f}_i , f_i$ represents the channel through which DM $\varphi$ decays into a pair of fermions.
\end{tablenotes}
\end{table}

\subsection{Branch Ratio}
\begin{table}
\centering
\caption{\label{decaymodes} Tree--level decay modes of the ssDM~\cite{Oscar}}
\begin{tabular}{|l |l|}
\hline
Decay mode
& Asymptotic scaling \\
\hline
$\varphi\to hh, WW, ZZ$
& $m_\varphi^3$ \\
% \hline
$\varphi\to f\bar{f}$
& $m_\varphi m_f^2$ \\
\hline
$\varphi\to hhh$
& $m_\varphi v^2$ \\
%	\hline
$\varphi\to WWh,ZZh$
& $m_\varphi^5/ v^2$ \\
%	\hline
$\varphi\to f\bar{f}h$
& $m_\varphi^3 m_f^2/ v^2$ \\
% \hline
$\varphi\to f'\bar{f}W,f\bar{f}Z$
& $m_\varphi^5/ v^2$ \\
% \hline
$\varphi\to ff\gamma,q\bar{q}g$
& $m_\varphi^3$ \\
\hline
$\varphi\to hhhh$
& $m_\varphi^3$ \\
% \hline
$\varphi\to WWhh,ZZhh$
& $m_\varphi^7/v^4$ \\
\hline
\end{tabular}
\end{table}

\begin{figure}[tbp]
\centering
\includegraphics[scale=0.51]{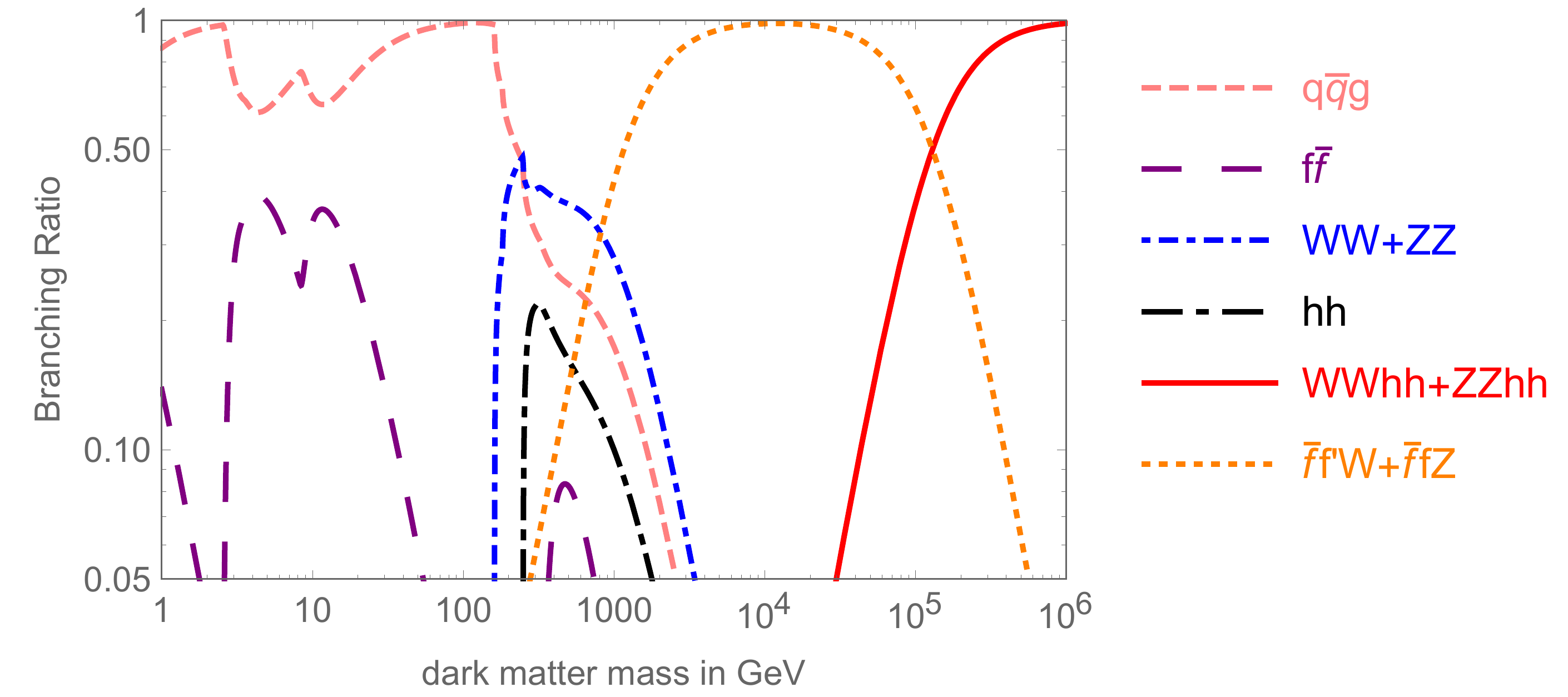}
\caption{\label{branchratio}Decay branch ratios of the ssDM via non-minimal coupling to gravity.
}
\end{figure}

Following the procedure provided by O. Cat\`{a} et al.~\cite{Oscar}, we draw decay branch ratios of the ssDM, which is shown in Fig.~\ref{branchratio}. And O. Cat\`{a} et al. also provided the asymptotic dependence of the corresponding partial width on the ssDM mass, in the limit of the massless final--state Standard Model particles, as shown in Table~\ref{decaymodes}.
This work focuses on the ssDM with a mass around the electroweak scale.

Below the electroweak scale ($m_\varphi < v$), decay branch ratio is dominated by $\varphi\to q\bar{q}g$ channel. Although the asymptotic scaling of $\varphi\to ff\gamma$ channel is also $m_\varphi^3$, this channel is suppressed by $\alpha_{em}/\alpha_s$. Compared with $\varphi\to q\bar{q}g$ channel, $\varphi\to f\bar{f}h$ channel is suppressed by $m_f^2/v^2$.
The ratio of $\varphi\to f\bar{f}$ channel to $\varphi\to q\bar{q}g$ channel is $m_f^2/m_\varphi^2$. Therefore, when the mass of fermions is close to that of ssDM, the contribution of $\varphi\to f\bar{f}$ channel can not be ignored. It is logical to recognize that in Fig.~\ref{branchratio}, final--state particles of the hump around 10 GeV of $\varphi\to f\bar{f}$ channel are mainly tau leptons, charm quarks and bottom quarks, and final--state particles of the peak near 500 GeV are mainly top quarks.

Above the electroweak scale ($4\pi v \lesssim m_\varphi \lesssim 10^5~\text{GeV}$), decay branch ratio is dominated by $\varphi\to f'\bar{f}W+f\bar{f}Z$ channel. Compared with $\varphi\to f'\bar{f}W+f\bar{f}Z$ channel, $\varphi\to q\bar{q}g$ channel is suppressed by factor $v^2/m_\varphi^2$. Similarly, $\varphi\to hhh$ channel is suppressed by factor $v^4/m_\varphi^4$. Although the asymptotic scaling of $\varphi\to WWh+ZZh$ channel is same as $\varphi\to f'\bar{f}W+f\bar{f}Z$ channel, it is suppressed by the smaller phase space.

Around the electroweak scale ($m_\varphi\sim v$), many channels have the asymptotic scaling of $m_\varphi^3$, including $\varphi\to WW+ZZ+hh+q\bar{q}g+\bar{f}f'W+f\bar{f}Z$. Since the mass of the top quark is also near the electroweak scale, the contribution from $\varphi\to f \bar{f}$ channel is also can not be neglected. So near the electroweak scale, the decay channels are the most abundant and worth a thorough analysis.

Only channels shown in Fig.~\ref{branchratio} were included in the following numerical calculations.

\section{Decay spectrum induced by global symmetry breaking}\label{decayspectrumGSB}

\subsection{Decay spectrum at production}

Tanabashi et al. (Particle Data Group)~\cite{particleDG} provided a detailed procedure to calculate decay rates and decay spectrum at production. These authors gave expressions for differential decay rates, e.g. Eq.~\eqref{TotalDecayRate}, relativistically invariant three-body phase space, e.g. Eq.~\eqref{xiangspace3}, and relativistically invariant four-body phase space, e.g. Eq.~\eqref{xiangspace4}.

For the convenience of description, in the following, we mark the three product particles arising from three-body decay as particle 1, particle 2 and particle 3. We also use nomenclature for the rest frame of particle $i$ and particle $j$ as $F_{ij}$.

The expression of the differential decay rate is
\begin{equation}
d\Gamma=\frac{1}{2m_\varphi}|\mathcal{M}|^2d\Phi^{(n)}(m_\varphi;p_1,...,p_n),
\label{TotalDecayRate}
\end{equation}
where $\Gamma$ is the decay rate of $\varphi$ in its rest frame, $m_\varphi$ is mass of the DMp, $\mathcal{M}$ is the invariant matrix element, $\Phi^{(n)}$ is the $n$-body phase space, and $p_i$ is the four momentum of terminal particle $i$. We also use the definitions $p_{ij}=p_i+p_j$, $m_{ij}^2=p_{ij}^2$, so that the element of three body phase space $d\Phi^{(3)}$ can be written as
\begin{equation}
d\Phi^{(3)}=\frac{1}{2\pi} dm_{12}^2
\frac{1}{16\pi^2} \frac{|\vec{p}_1^*|}{m_{12}}d\Omega_{1}^*
\frac{1}{16\pi^2} \frac{|\vec{p}_3|}{m_\varphi}d\Omega_{3},
\label{xiangspace3}
\end{equation}
where($|\vec{p}_1^*|,\Omega_{1}^*$) is the three momentum of particle 1 in $F_{12}$, and $\Omega_3$ is the angle of particle 3 in the rest frame of the decaying particle. The symbol $*$ always denotes the quantity in $F_{12}$.

The relationship between $E_3$ and $m_{12}$ is
\begin{equation}
E_3=\frac{m_{\varphi}^2+m_{3}^2-m_{12}^2}{2m_{\varphi}},
\label{e3andm12}
\end{equation}
where $m_3$ and $E_3$ are the mass and energy of particle 3, respectively. The energy spectrum of particle 3 per decay in a channel with final state $l$ can be calculated following
\begin{equation}
\frac{d{N}^l}{dE_3}=
\frac{\partial\Gamma^l}{\Gamma^l \partial E_3}.
\label{definationdNdE}
\end{equation}

Using the Feynman rules given in Table~\ref{vertexrules}, and following Eqs.~\eqref{TotalDecayRate}, \eqref{xiangspace3}, \eqref{e3andm12} and \eqref{definationdNdE}, we numerically calculated the decay rate $\Gamma$ and energy spectrum $d{N}^l/dE_3$. According to translatable symmetry, $d{N}^l/dE_1$ and $d{N}^l/dE_2$ were also calculated, where $E_1$ is the energy of particle 1, $E_2$ is the energy of particle 2.

As for four-body decay, there are three channels: $\varphi\to W^+,W^-,h,h$; $\varphi\to Z,Z,h,h$ and $\varphi\to h,h,h,h$. We will consider $\varphi\to W^+,W^-,h,h$ here to illustrate our method of calculation.
To demonstrate the calculation of $\Gamma$ and $d{N}^l/dE_1$ clearly, we regard the $W^+$ boson as particle 1 and the $W^-$ boson as particle 2, while the remaining two Higgs bosons are particles 3 and 4. As before, we still denote the rest frame of particles $i$ and $j$ as $F_{ij}$.

The element of four-body phase space $d\Phi^{(4)}$ can be written as
\begin{equation}
d\Phi^{(4)}=
\frac{1}{2\pi} dm_{12}^2
\frac{1}{2\pi} dm_{34}^2
\frac{1}{16\pi^2} \frac{|\vec{p}_1^*|}{m_{12}}d\Omega_{1}^*
\frac{1}{16\pi^2} \frac{|\vec{p}_3^{**}|}{m_{34}}d\Omega_{3}^{**}
\frac{1}{16\pi^2} \frac{|\vec{p}_{12}|}{m_\varphi}d\Omega_{12},
\label{xiangspace4}
\end{equation}
where($|\vec{p}_{12}|,\Omega_{12}$) is the three momentum of $p_{12}$, and $(\vec{p}_3^{**},\Omega_{3}^{**})$ is the three momentum of particle 3 in $F_{34}$. The symbol $**$ always denotes the quantity in $F_{34}$. Using Eqs.~\eqref{TotalDecayRate} and \eqref{xiangspace4}, we numerically calculated $\Gamma$ and $\partial^2{N}^l/(\partial m_{12}\partial m_{34})$, where $\partial^2{N}^l/(\partial m_{12}\partial m_{34})=\partial^2{\Gamma}^l/(\Gamma\partial m_{12}\partial m_{34})$. Then we applied Lorentz transformations to $|\vec{p}_1^*|$ and $E_1^*$. We find that the isotropic spectrum of particle 1 with momentum $|\vec{p}_1^*|$ in $F_{12}$ has a spectrum described by Eq.~\eqref{spectrumform} in the rest frame of $\varphi$,
\begin{equation}
g(E_1,m_{12})=\frac{1}{2}\frac{1}{\gamma_{12}\beta_{12}|\vec{p}_1^*|} \Theta(E_1-E_-) \Theta(E_+-E_1)
\label{spectrumform}
\end{equation}
where $\beta_{ij}$ is the velocity of $F_{ij}$ relative to the decaying DMp, $\gamma_{ij}=(1-\beta_{ij}^2)^{-1/2}$, $E_\pm \equiv \gamma_{12}E_1^* \pm \gamma_{12}\beta_{12}|\vec{p}_1^*|$ and $\Theta(x)$ the Heaviside function.

The energy spectrum of particle 1 produced per decay in the channel with final state $l$ can be described by
\begin{equation}
\frac{d{N}^l}{dE_1}=\int\int
g(E_1,m_{12})
\frac{\partial^2{N}^l}{\partial m_{12}\partial m_{34}}
dm_{12}dm_{34}.
\label{fourbodydNdE}
\end{equation}

As before, according to translatable symmetry, $d{N}^l/dE_2$, $d{N}^l/dE_3$ and $d{N}^l/dE_4$ can also be calculated, where $E_2$, $E_3$ and $E_4$ represent the energy of particles 2, 3 and 4 respectively.

So far, we have obtained many spectra of stable and unstable particles, such as of the Higgs boson, Z boson and neutrino. For comparison with observations, we should further calculate the spectra of the final--state stable particles, specifically, photons and positrons. Cirelli et al.~\cite{cookbook} use the \textsc{Pythia} codes to generate spectra of photons and positrons $k(E,E_{\gamma,e^+})$ induced by a primary state particle with given energy $E$, where $E_{\gamma,e^+}$ represents energy of the photon or positron. The effect of QED and EW Bremsstrahlung are included when they used \textsc{Pythia} to generate $k(E,E_{\gamma,e^+})$, while the effects of Inverse Compton processes or synchrotron radiation are not included~\cite{cookbook}. Then, the secondary photon or positron energy spectrum produced per decay in a channel with final state $l$ represented by $d{N}^l/dE_{\gamma,e^+}$ was numerically calculated as
\begin{equation}
\frac{d{N}^l}{dE_{\gamma,e^+}}
=\sum_s
\int k(E_s,E_{\gamma,e^+}) \frac{d{N}^l}{dE_s}
dE_s
\label{secondaryspectrum}
,
\end{equation}
where $s$ includes all final state particles in the channel with final state $l$. In the three-body decay case, $s$ runs from 1 to 3, while in the four-body decay case $s$ runs from 1 to 4.

\subsection{Fluxes after propagation}

Finally, the spectra that could be detected by satellites are calculated via PPPC 4 DM ID~\cite{cookbook}. In the following, we uniformly adopt the Navarro-Frenk-White (NFW) DM distribution model
\begin{equation}
\rho(r)=\rho_s\frac{r_s}{r}(1+\frac{r}{r_s})^{-2}
\end{equation}
with parameters $\rho_s=0.184~\text{GeV}/\text{cm}^3$, $r_s=24.42~\text{kpc}$, where $\rho(r)$ is the energy density of DM at a distance of $r$ from the Galactic Center.

The differential flux of positrons in space $\vec{x}$ and time $t$ is given by $d\Phi_{e^+}/dE_{e^+}(t,\vec{x},E_{e^+})=v_{e^+}f/4\pi$, where $v_{e^+}$ is the velocity of the positrons. The positron number density per unit energy $f$ obeys the diffusion--loss equation~\cite{cookbook}~\cite{galacticPositron}
\begin{equation}
\frac{\partial f}{\partial t}-\triangledown(\mathcal{K}(E_{e^+},\vec{x})\triangledown f)-\frac{\partial}{\partial E_{e^+}}(b(E_{e^+},\vec{x})f)=Q(E_{e^+},\vec{x})
,
\label{diffeq}
\end{equation}
where $\mathcal{K}(E_{e^+},\vec{x})$ is the diffusion coefficient function which describe transport through the turbulent magnetic fields.
We adopt the customary parameterization $\mathcal{K}=\mathcal{K}_0(E_{e^+}/\text{GeV})^\delta=\mathcal{K}_0 \epsilon^\delta$ with the parameters $\mathcal{K}_0=0.0112~\text{kpc}^2/\text{Myr}$ and $\delta=0.70$, which would result in a median final result~\cite{cookbook}. $b(E_{e^+},\vec{x})$ is the energy loss coefficient function which describes the energy loss due to several processes, such as synchrotron radiation and Inverse Compton scattering (ICS) off CMB photons, and/or infrared and optical galactic starlight, it is provided numerically by PPPC 4 DM ID~\cite{cookbook} in the form of \textsc{Mathematica}\textsuperscript{\textregistered} interpolating functions.
$Q$ is the source term which can be expressed as
\begin{equation}
Q=\frac{\rho(r)}{m_\varphi}\sum_l \Gamma_l \frac{dN_{e^+}^l}{dE_{e^+}}.
\end{equation}
Eq.~\ref{diffeq} is solved in a cylinder that sandwiches the galactic plane with height $2L$ and radius $R=20~\text{kpc}$. The distance between the solar system and the Galactic Center is 8.33 kpc. Conditions electrons/positrons could escape freely are adopted on the surface of the cylinder. The resulting differential flux of positrons in the Solar System is
\begin{equation}
\frac{d\Phi_{e^+}}{dE_{e^+}}(E_{e^+},r_\odot)=
\frac{v_{e^+}}{4\pi b(E_{e^+},r_\odot)}
\frac{\rho_\odot}{m_\varphi} \sum_l \Gamma_l
\int_{E_{e^+}}^{m_\varphi/2}
dE_s \frac{dN^l_{e^+}}{dE_{e^+}}(E_s) I(E_{e^+},E_s,r_\odot)
\end{equation}
where $r_\odot$ is the distance between the Solar System and the Galactic Center, and $\rho_\odot$ is the DM density at the Solar System. $E_s$ is the positron energy at production ($s$ stands for "source"), $I(E_{e^+},E_s,r_\odot)$ is the generalized halo function, which is the Green function from a source with positron energy $E_s$ to any energy $E_{e^+}$, and it is also provided numerically by PPPC 4 DM ID~\cite{cookbook} in the form of \textsc{Mathematica}\textsuperscript{\textregistered} interpolating functions.

The calculation of gamma rays consists of three parts, direct ("prompt") decay from the Milky Way halo, extragalactic gamma rays emitted by DM decay and gamma rays from Inverse Compton scattering (ICS). The synchrotron radiation is in a significant amount from where the magnetic field and the DM are very dense, close to the Galactic Center. This work focuses on high galactic latitude ($|b|>20^\circ$), where the magnetic field is very weak, so synchrotron radiation is not included in this work.

The differential flux of photons from prompt decay of the Milky Way halo is calculated via
\begin{equation}
\frac{d\Phi_\gamma}{dE_\gamma d\Omega}=\frac{r_\odot \rho_\odot}{4\pi m_\varphi} \bar{J} \sum_l \Gamma_l \frac{dN^l_\gamma}{dE_\gamma}
\end{equation}
where $\bar{J}(\triangle\Omega)=\int_{\triangle\Omega} J d\Omega/\triangle\Omega$ is the averaged $J$ factor of the region of interest, $J=\int_{\text{l.o.s.}} \rho(r(s,\theta))/(r_\odot\rho_\odot) ds$, $r(s,\theta)=(r_\odot^2+s^2-2 r_\odot s \text{cos}\theta)^{1/2}$ is the distance between the DM and the Galactic Center, and $\theta$ is the angle between the direction of the line of sight (l.o.s.) and the line connecting the Sun to the Galactic Center.

The extragalactic gamma rays received at a point with redshift $z$ is calculated via~\cite{cookbook}
\begin{equation}
\frac{d\Phi_{\text{EG}\gamma}}{dE_\gamma}(E_\gamma,z)=
\frac{c}{E_\gamma}\int_{z}^{\infty} dz' \frac{1}{H(z')(1+z')}(\frac{1+z}{1+z'})^3
\frac{1}{4 \pi}
\frac{\bar{\rho}(z')}{m_\varphi}\sum_l \Gamma_l \frac{dN^l_\gamma}{dE_\gamma'}(E_\gamma')
e^{-\tau(E_\gamma',z,z')}
\end{equation}
where $H(z)=H_0\sqrt{\Omega_m (1+z)^3+(1-\Omega_m)}$ is the Hubble function, $\bar{\rho}(z)=\bar{\rho}_0(1+z)^3$ is the average cosmological DM density and $\bar{\rho}_0 \simeq 1.15\times 10^{-6}~\text{GeV}/\text{cm}^3$, $E_\gamma'=E_\gamma(1+z')$, $\tau(E_\gamma',z,z')$ is the optical depth, which is also provided numerically by PPPC 4 DM ID~\cite{cookbook} in the form of \textsc{Mathematica}\textsuperscript{\textregistered} interpolating functions. $\tau(E_\gamma',z,z')$ describes the absorption of gamma rays in the intergalactic medium between the redshifts $z$ and $z'$. The presence of ultraviolet (UV) background lower the UV photon densities. There are three absorption models provided by PPPC 4 DM ID~\cite{cookbook}, (no ultraviolet (noUV), minimal ultraviolet (minUV) and maximal ultraviolet (maxUV)). We calculated the Hubble function in the $\Lambda$CDM cosmology with a pressure-less matter density of the universe $\Omega_m=0.27$, dark energy density of the universe $\Omega_\Lambda=0.73$ and scale factor for Hubble expansion rate $0.7$.

Galactic electrons/positrons generated by ssDM could lose their energy into photons by means of the inverse Compton scattering. The greater the mass of ssDM is, the higher the energy of the electrons/positrons generated by ssDM is, and the more important this effect is. The inverse Compton gamma rays is calculated as follows,
\begin{equation}
\frac{d\Phi_{\text{IC}\gamma}}{dE_\gamma d\Omega}=
\frac{1}{E_\gamma^2}\frac{r_\odot}{4\pi}\frac{\rho_\odot}{m_\varphi}
\int_{m_e}^{m_\varphi/2} dE_s \sum_i \Gamma_i \frac{dN_{e^+}^i}{dE}(E_s) I_{\text{IC}}(E_\gamma,E_s,b,l)
\end{equation}
where $b$ and $l$ is the galactic latitude and galactic longitude respectively, $I_{\text{IC}}(E_\gamma,E_s,b,l)$ is a halo function for the IC radiative process, which is also provided numerically by PPPC 4 DM ID~\cite{cookbook} in the form of \textsc{Mathematica}\textsuperscript{\textregistered} interpolating functions.

\section{Constraints from isotropic diffuse $\gamma$-ray background (IGRB) and the cosmic positron spectrum}\label{ConstraintsfromAMSandFermiLAT}

\subsection{Statistic methods to set constraints}

The isotropic diffuse $\gamma$-ray background is measured by Fermi-LAT~\cite{FermiLAT}. We compared the $\gamma$-ray flux produced by DM with IGRB to set constraints on the lifetime of ssDM. The region of interest in our work only includes high-latitude regions ($|b|>20^\circ$) because the analysis of the IGRB by Fermi-LAT only includes high-latitude ($|b|>20^\circ$)~\cite{FermiLAT}, where $b$ is the galactic latitude.

The cosmic positron flux is measured by the AMS on the International Space Station~\cite{AMS02}. We also compared the positron flux produced by DM with the measured flux to set constraints on the lifetime of ssDM.

The comparison strategies used in this paper are as follows. Define $\chi^2$ as
\begin{equation}
\chi^2=\sum_i \frac{(\Phi^{\text{th}}_i-\Phi^{\text{obs}}_i)^2}{\delta_i^2}
\Theta(\Phi^{\text{th}}_i-\Phi^{\text{obs}}_i)
\end{equation}
where $\Phi^{\text{th}}_i$ and $\Phi^{\text{obs}}_i$ denote the predicted and observed fluxes respectively, $\delta_i$ are the experimental errors, and $\Theta(x)$ is the Heaviside function. This work require $\chi^2<9$ to obtain an approximate estimate of 3-$\sigma$ constraint~\cite{chi2test1}~\cite{chi2test2} and only energy bins located at above 1 GeV are used.

\subsection{Treatment of the background}

\begin{figure}[htbp]
\subfigure[$m_\varphi=246~\text{GeV}$]{\includegraphics[scale=0.28]{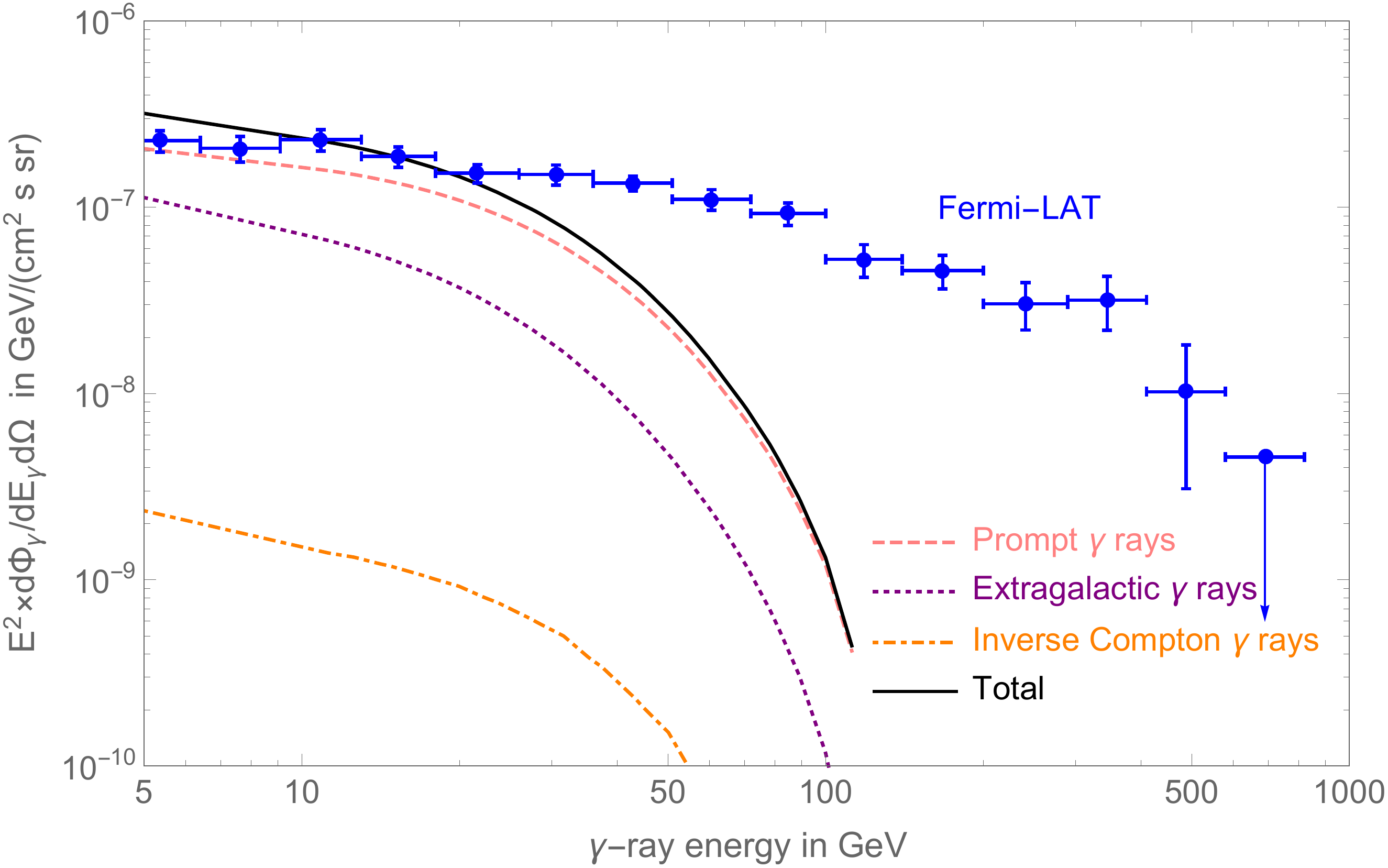}}
\hfill
\subfigure[$m_\varphi=500~\text{GeV}$]{\includegraphics[scale=0.3140]{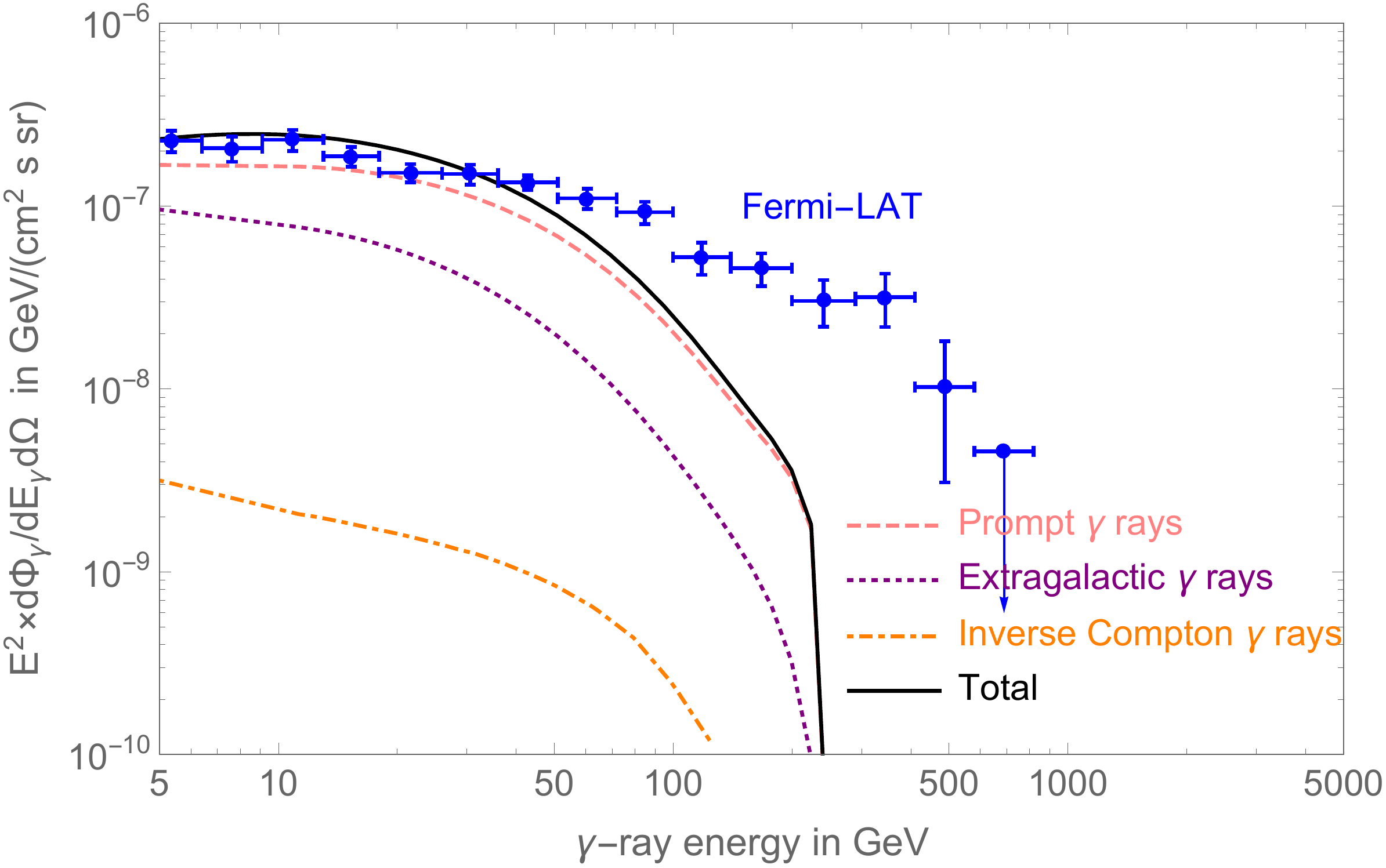}}
\hfill
\subfigure[$m_\varphi=1000~\text{GeV}$]{\includegraphics[scale=0.28]{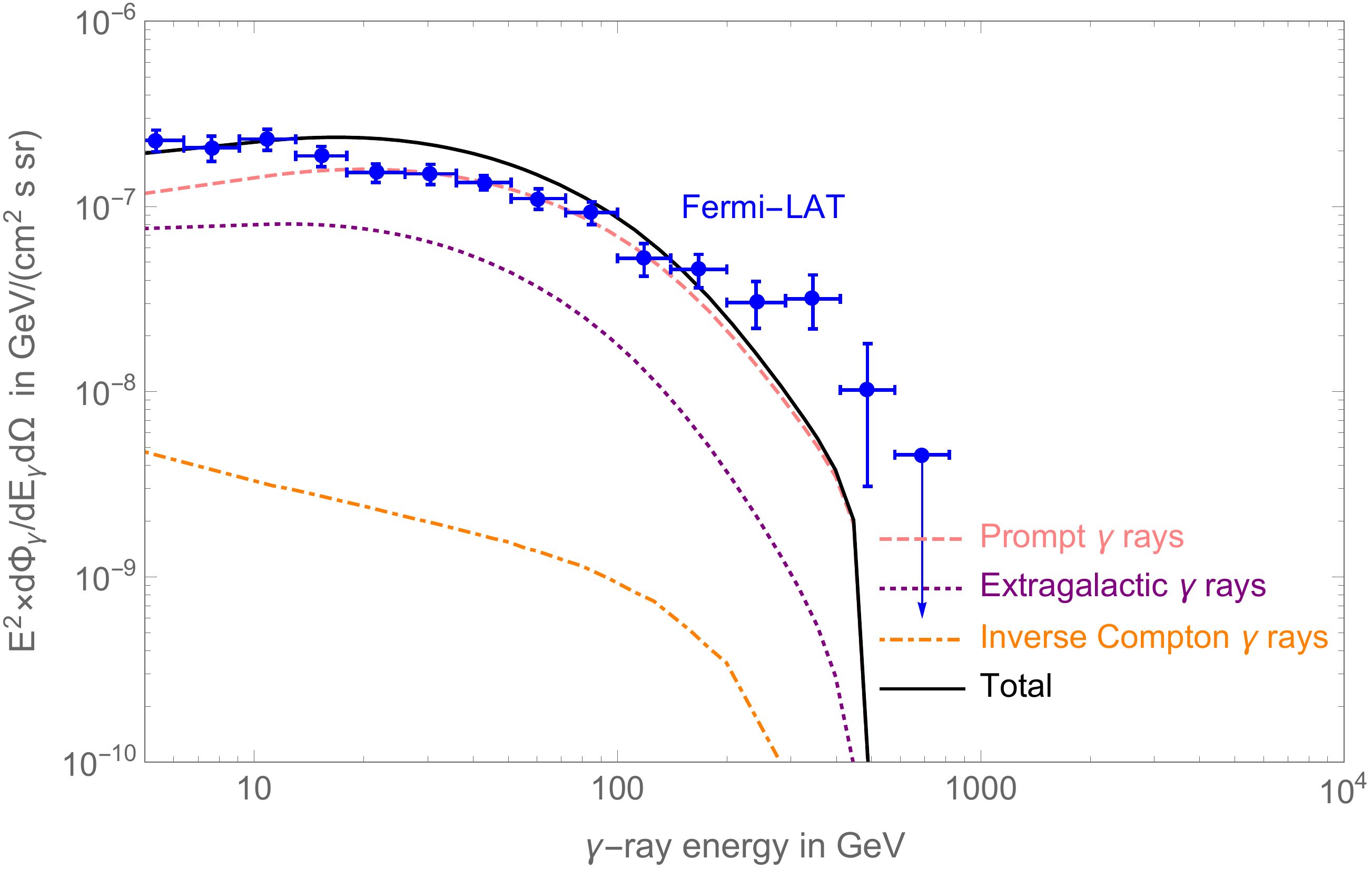}}
\hfill
\subfigure[$m_\varphi=20~\text{TeV}$]{\includegraphics[scale=0.328]{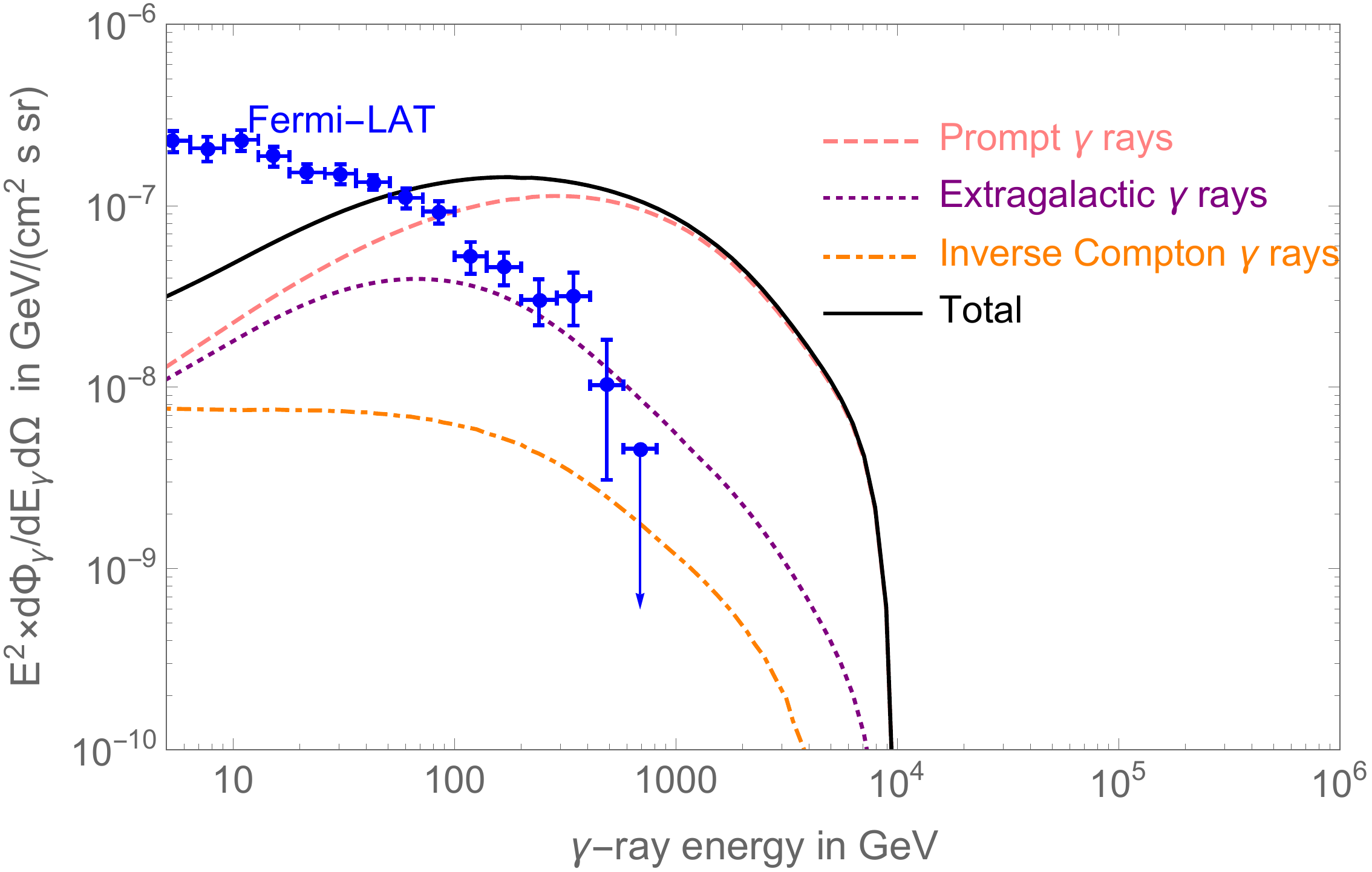}}
\caption{\label{1gammaflux}
Averaged photon fluxes ($|b|>20^\circ$) from decaying DMps of prompt emission component, extragalactic component, and inverse Compton scattering component are shown by the dashed line, dotted line and dot--dashed line respectively when $\tau=5.3\times10^{26}~s$. The total flux of these three components is shown by the black solid line. Fermi-LAT observations of the IGRB are also plotted by blue points with error bars.
}
\end{figure}
\begin{figure}[htbp]
\subfigure[$m_\varphi=246~\text{GeV}$]{\includegraphics[scale=0.335]{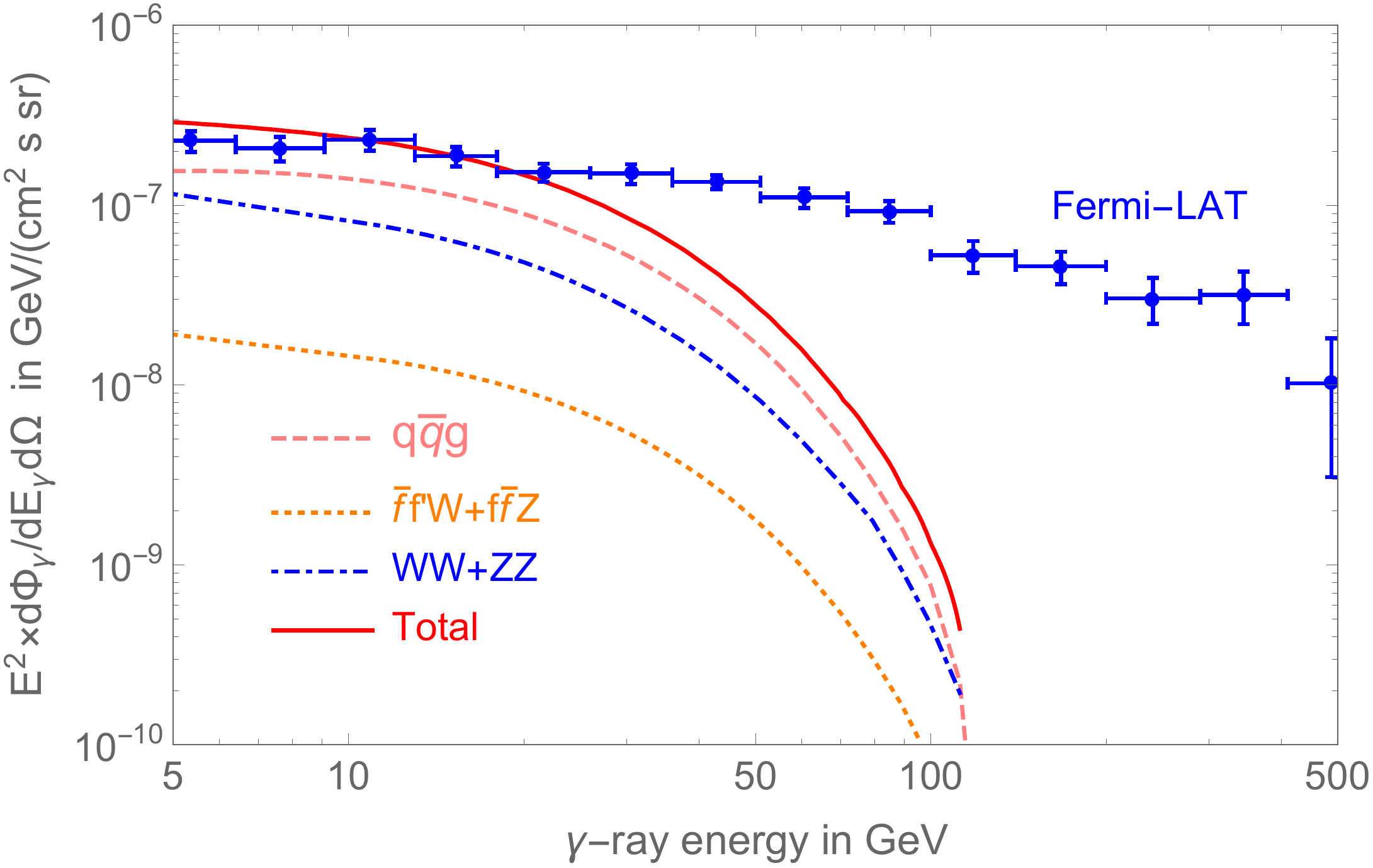}}
\hfill
\subfigure[$m_\varphi=500~\text{GeV}$]{\includegraphics[scale=0.32]{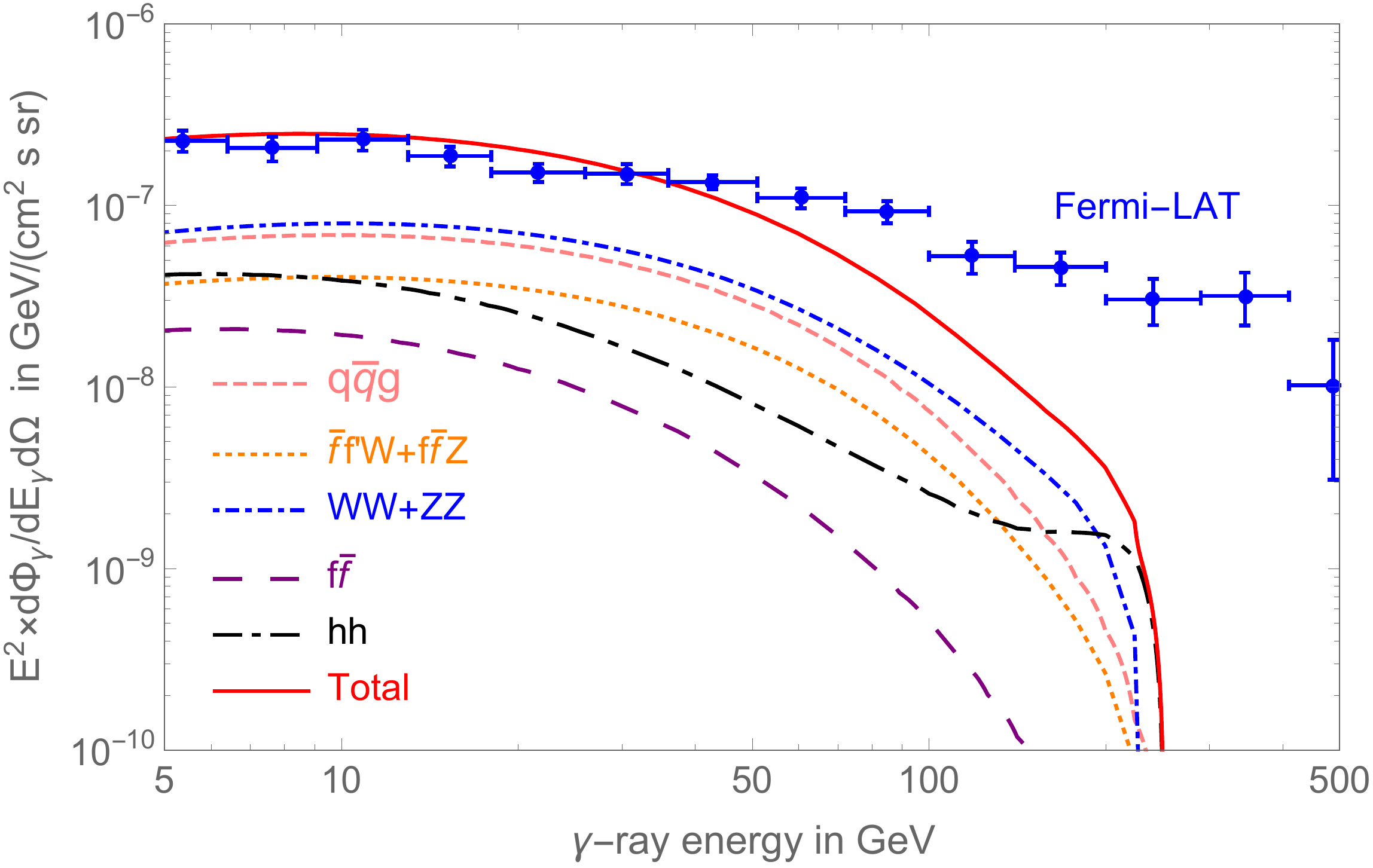}}
\hfill
\subfigure[$m_\varphi=1000~\text{GeV}$]{\includegraphics[scale=0.31]{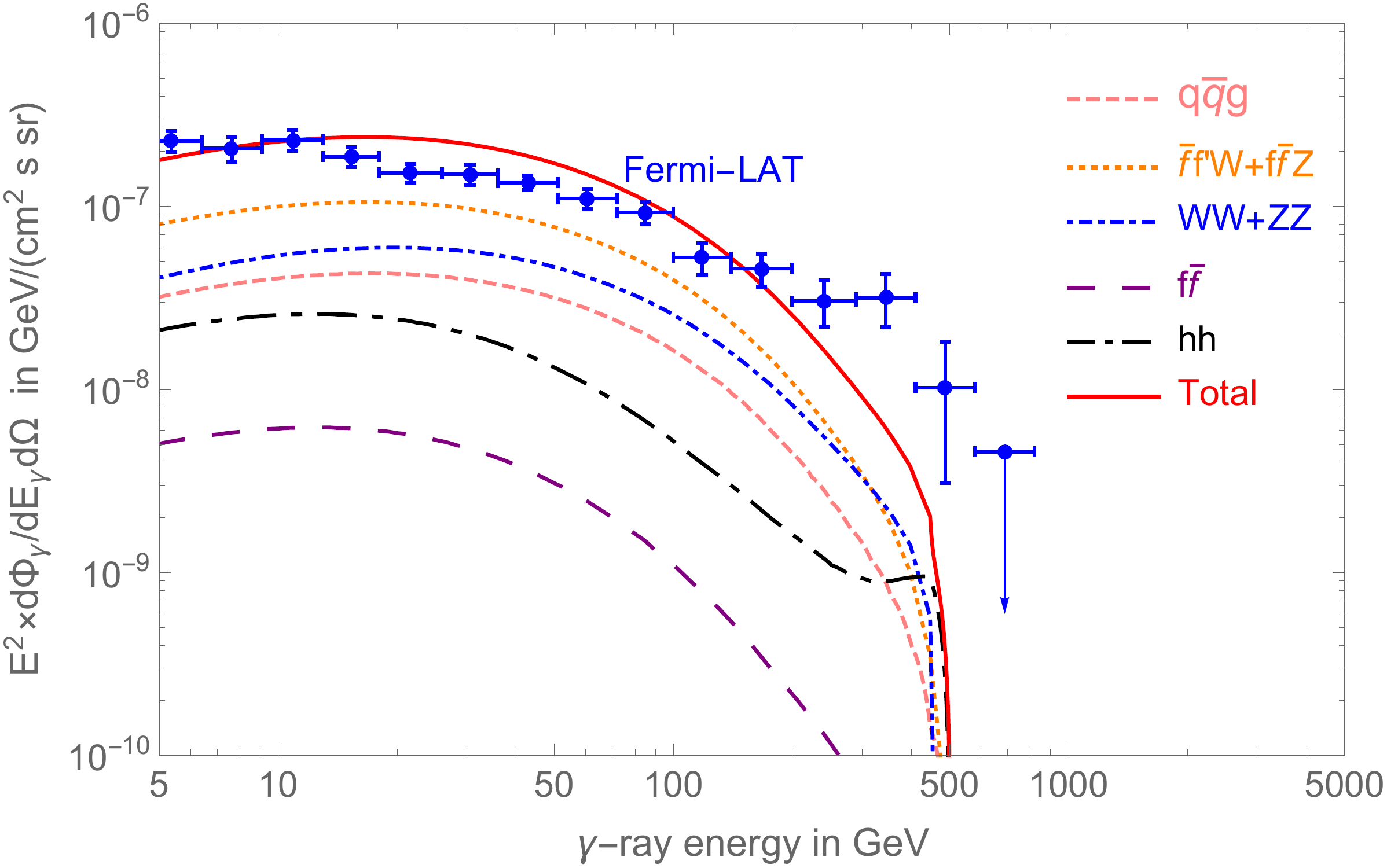}}
\hfill
\subfigure[$m_\varphi=20~\text{TeV}$]{\includegraphics[scale=0.343]{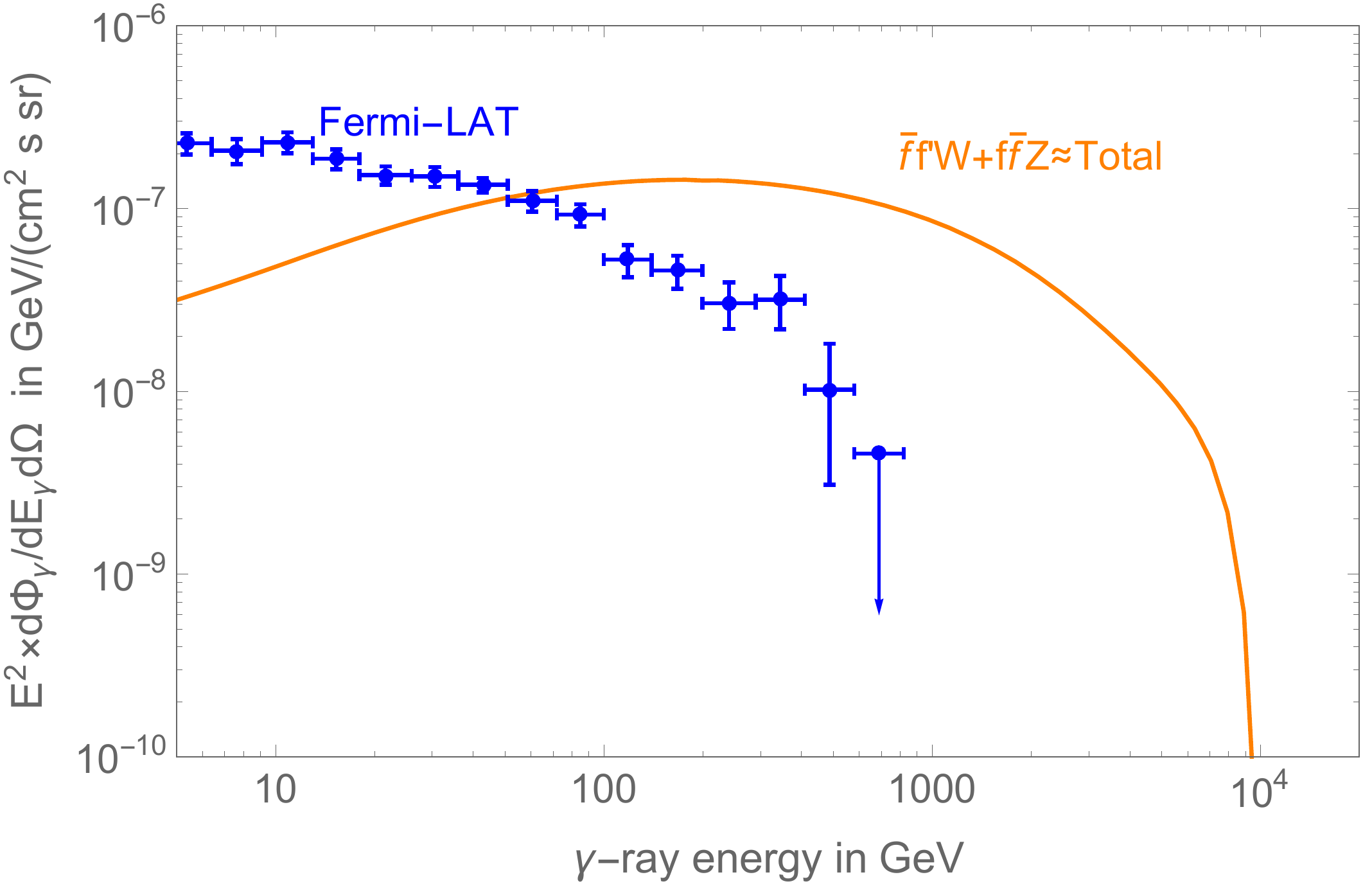}}
\caption{\label{2gammaflux}
Averaged photon flux ($|b|>20^\circ$) from decaying DMps contributed by various channels are shown when $\tau=5.3\times10^{26}~s$. The total flux is shown by the solid line. Fermi-LAT observations of the IGRB are also shown by blue points with error bars.
}
\end{figure}
\begin{figure}[htbp]
\subfigure[$m_\varphi=246~\text{GeV}$]{\includegraphics[scale=0.4]{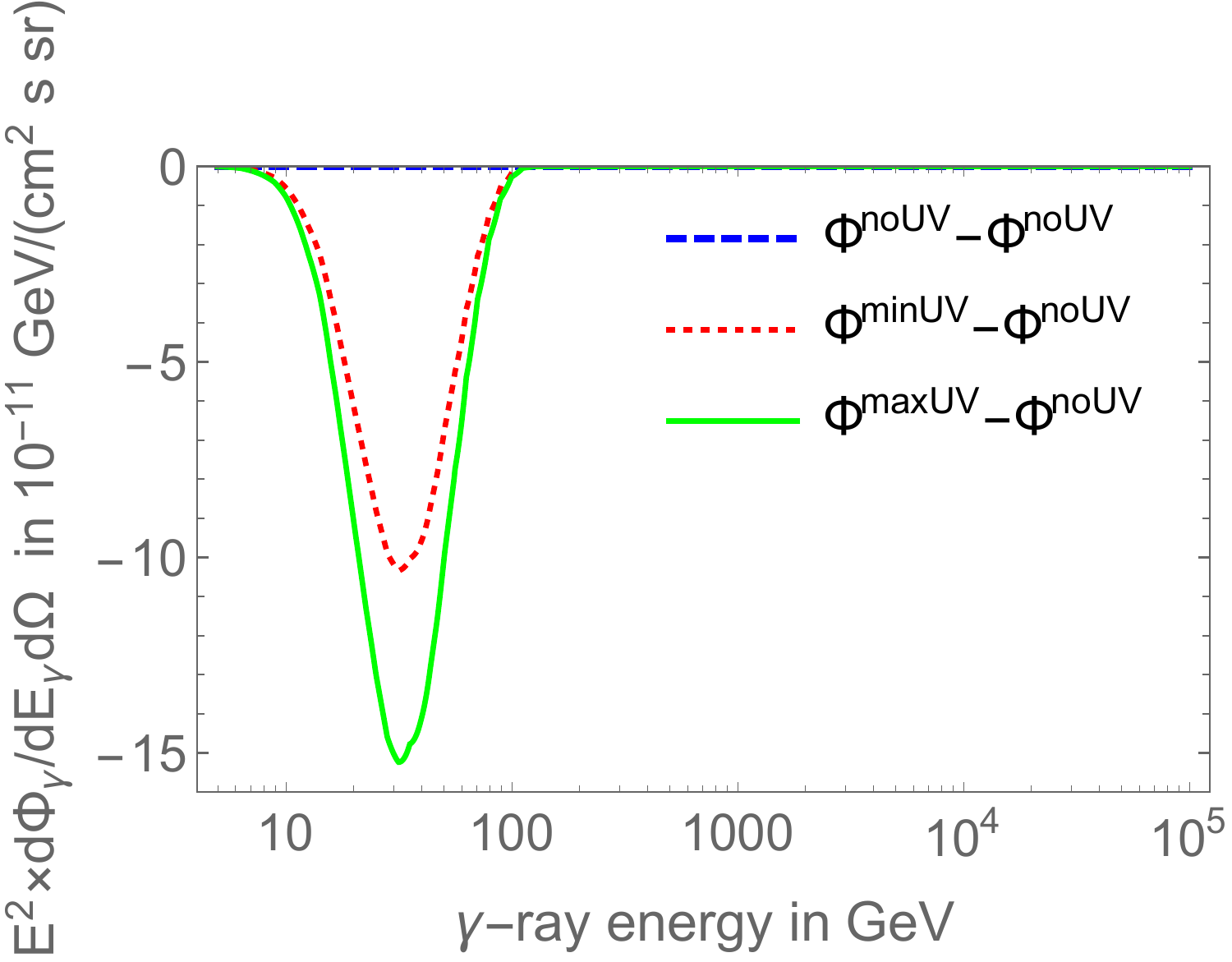}}
\hfill
\subfigure[$m_\varphi=500~\text{GeV}$]{\includegraphics[scale=0.4]{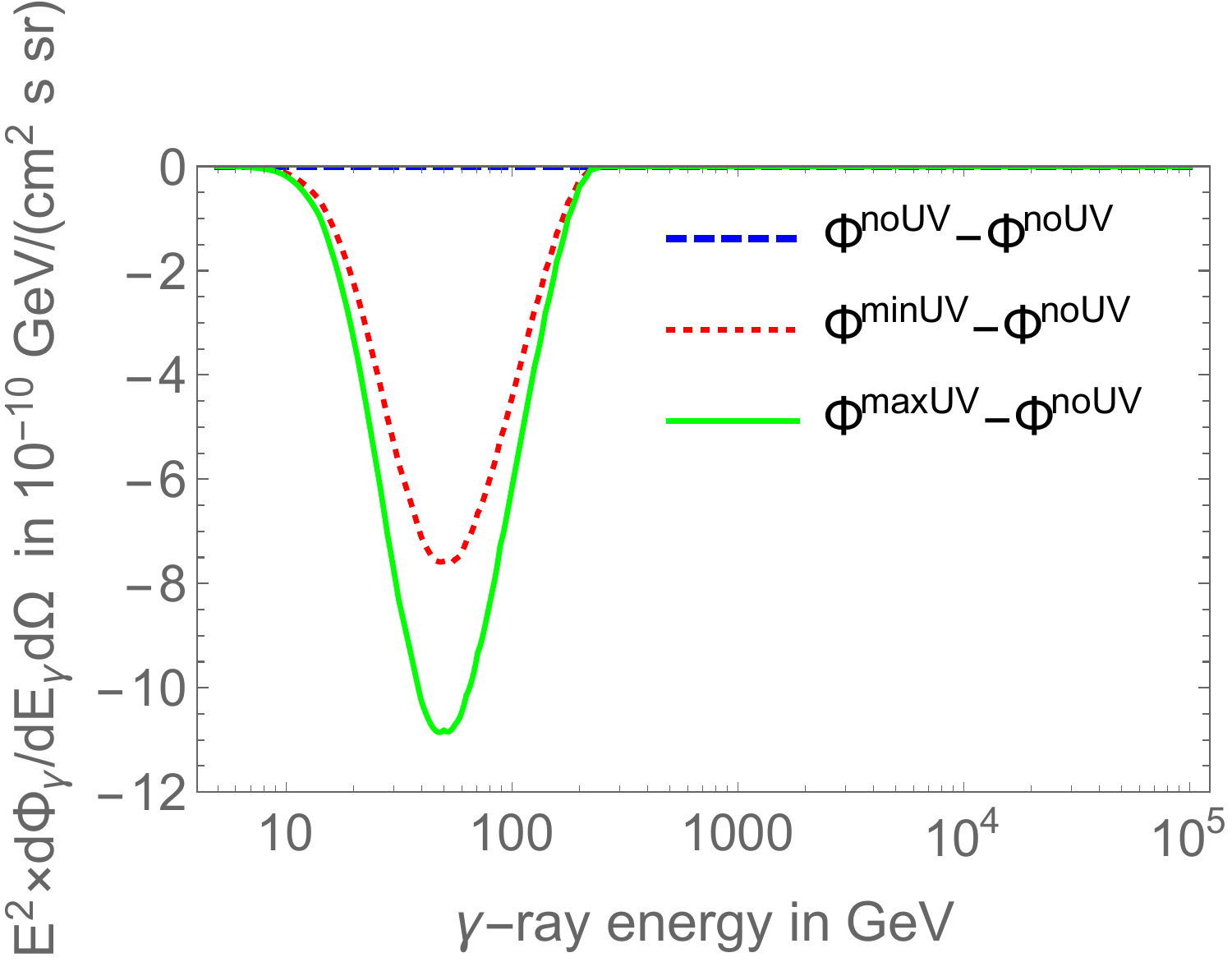}}
\hfill
\subfigure[$m_\varphi=1000~\text{GeV}$]{\includegraphics[scale=0.4]{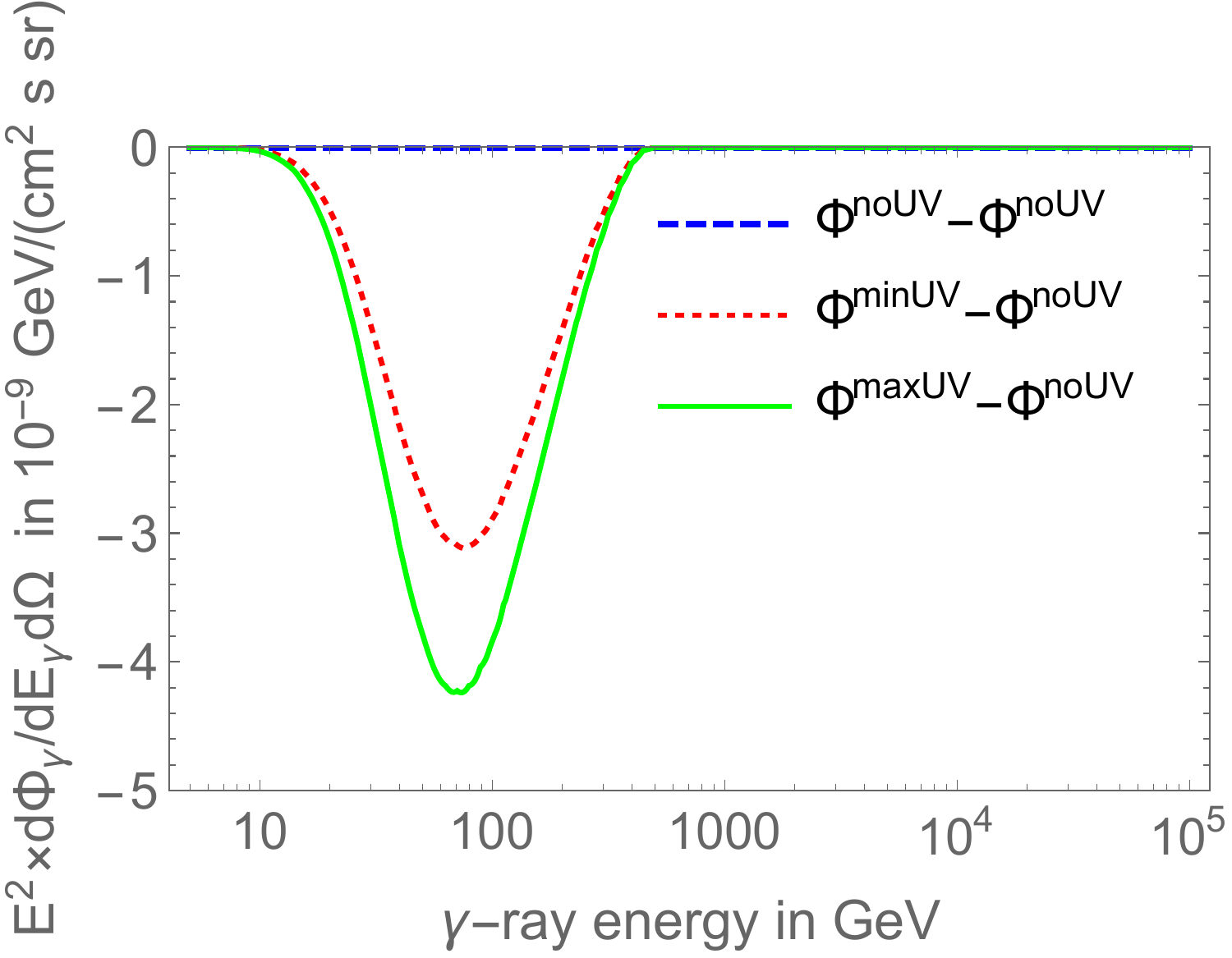}}
\hfill
\subfigure[$m_\varphi=20~\text{TeV}$]{\includegraphics[scale=0.4]{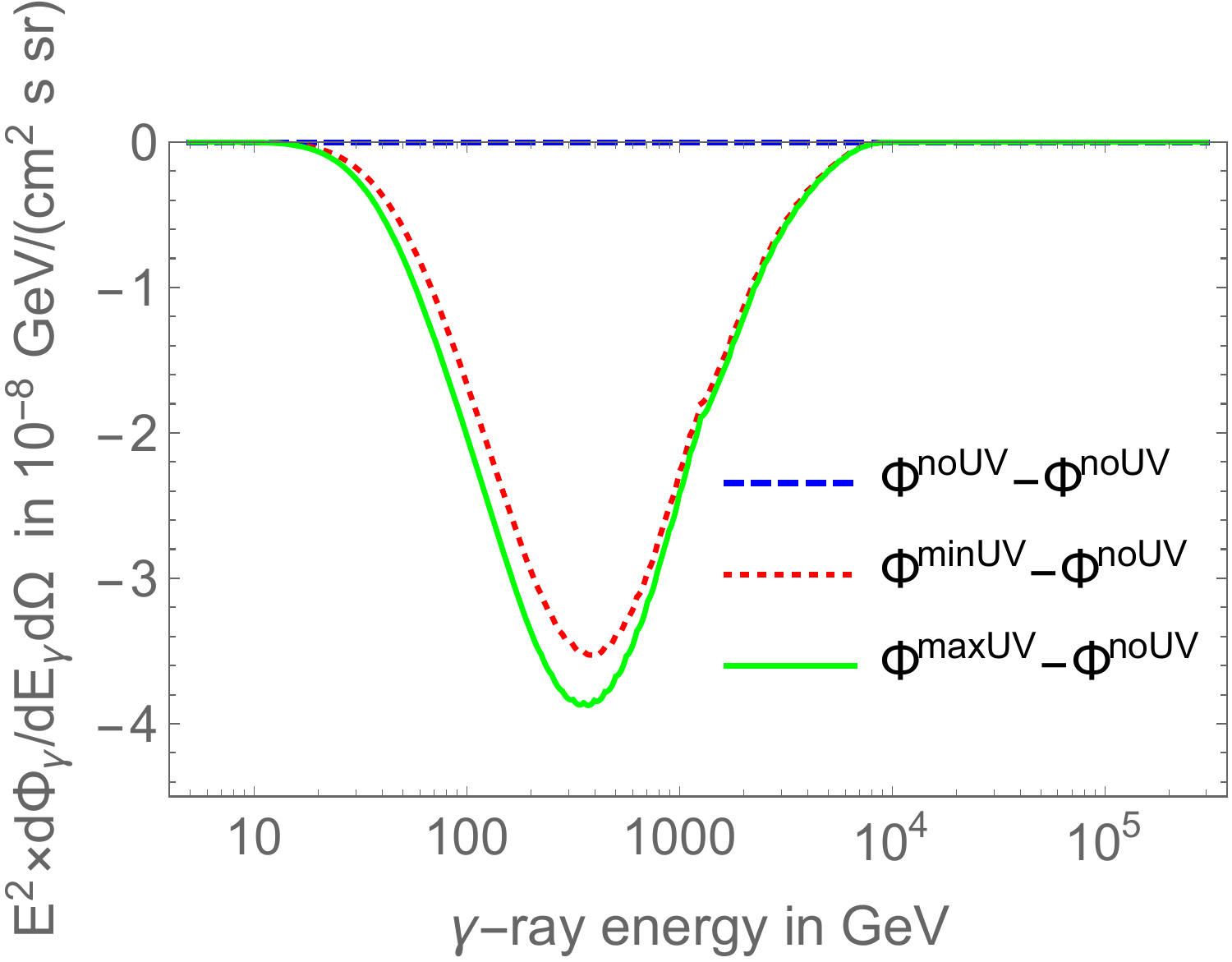}}
\caption{\label{3gammaflux}
The presence of UV background lower the UV photon densities. The figure shows the absorption of UV photons compared with no UV background regime when $\tau=5.3\times10^{26}~s$.
}
\end{figure}
\begin{figure}[htbp]
\subfigure[$m_\varphi=246~\text{GeV}$]{\includegraphics[scale=0.387]{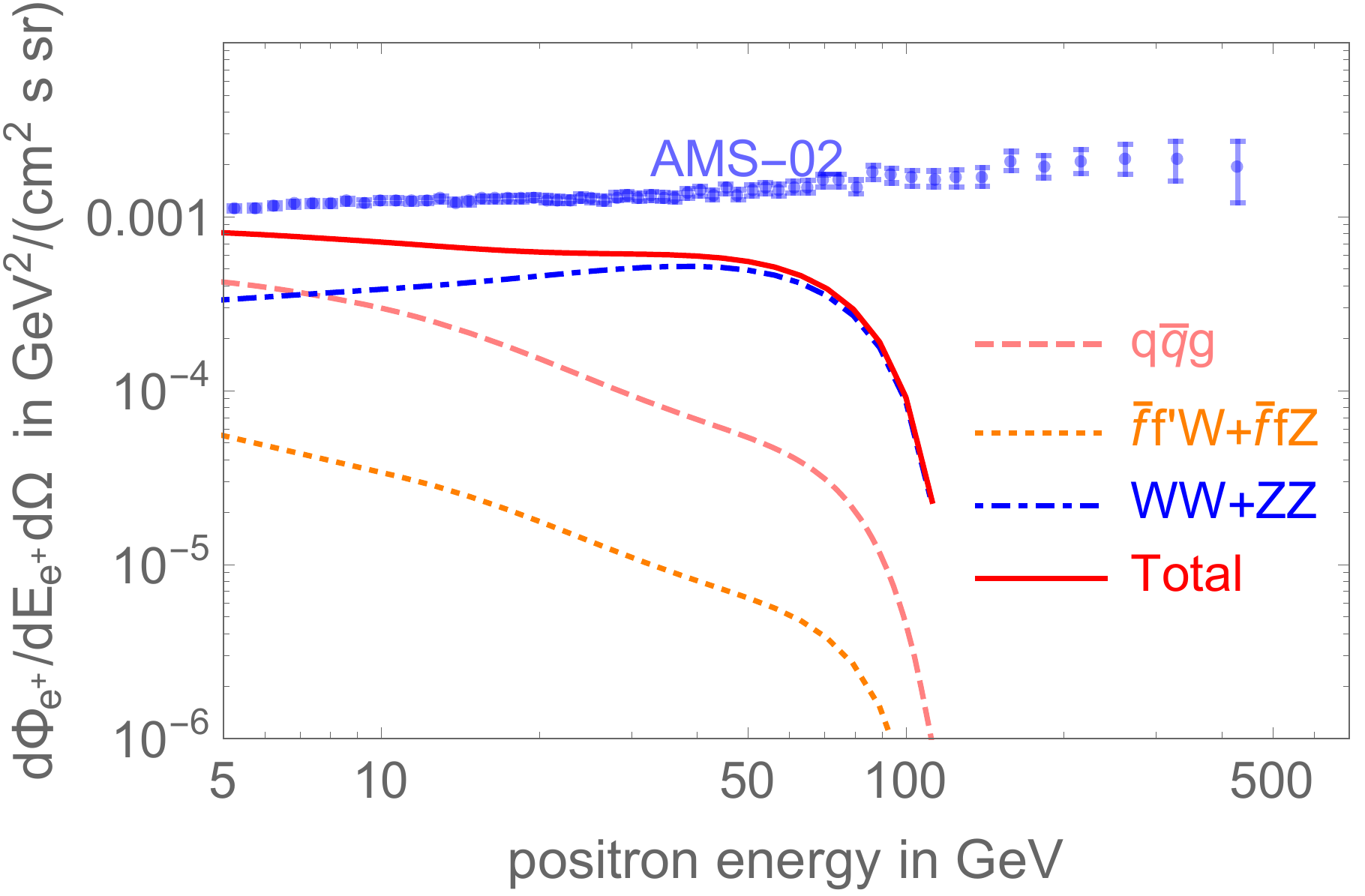}}
\hfill
\subfigure[$m_\varphi=500~\text{GeV}$]{\includegraphics[scale=0.352]{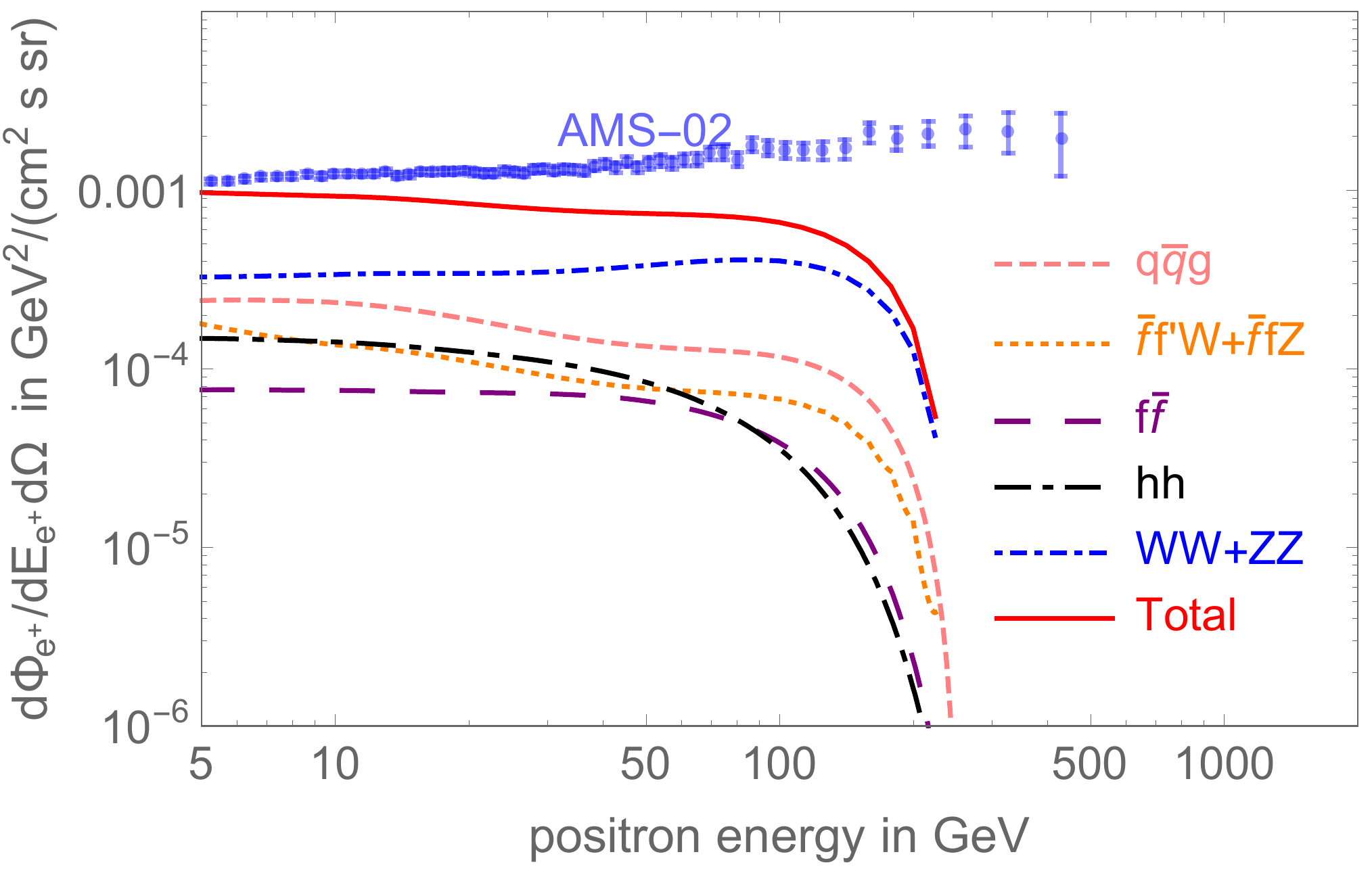}}
\hfill
\subfigure[$m_\varphi=1000~\text{GeV}$]{\includegraphics[scale=0.36]{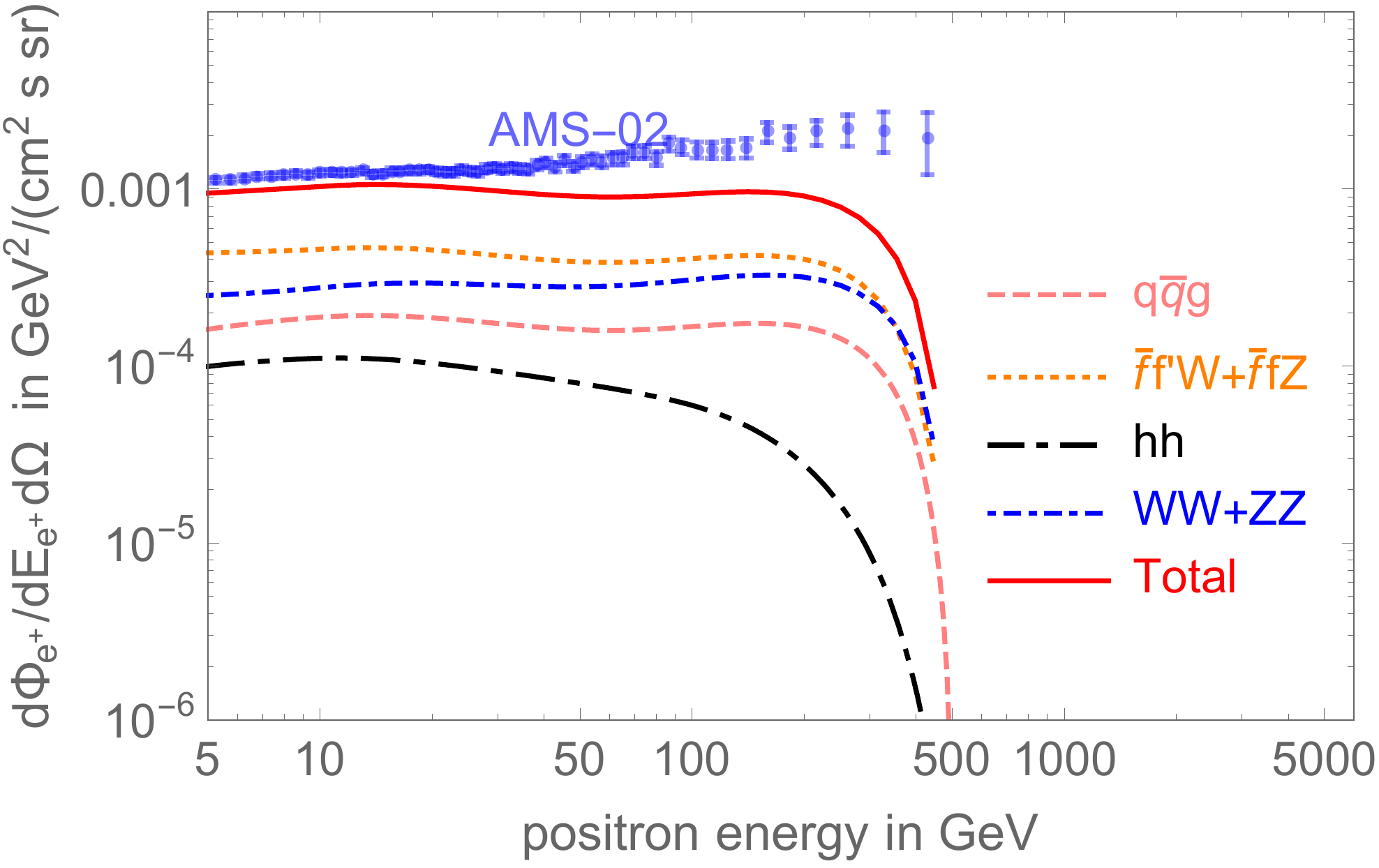}}
\hfill
\subfigure[$m_\varphi=20~\text{TeV}$]{\includegraphics[scale=0.362]{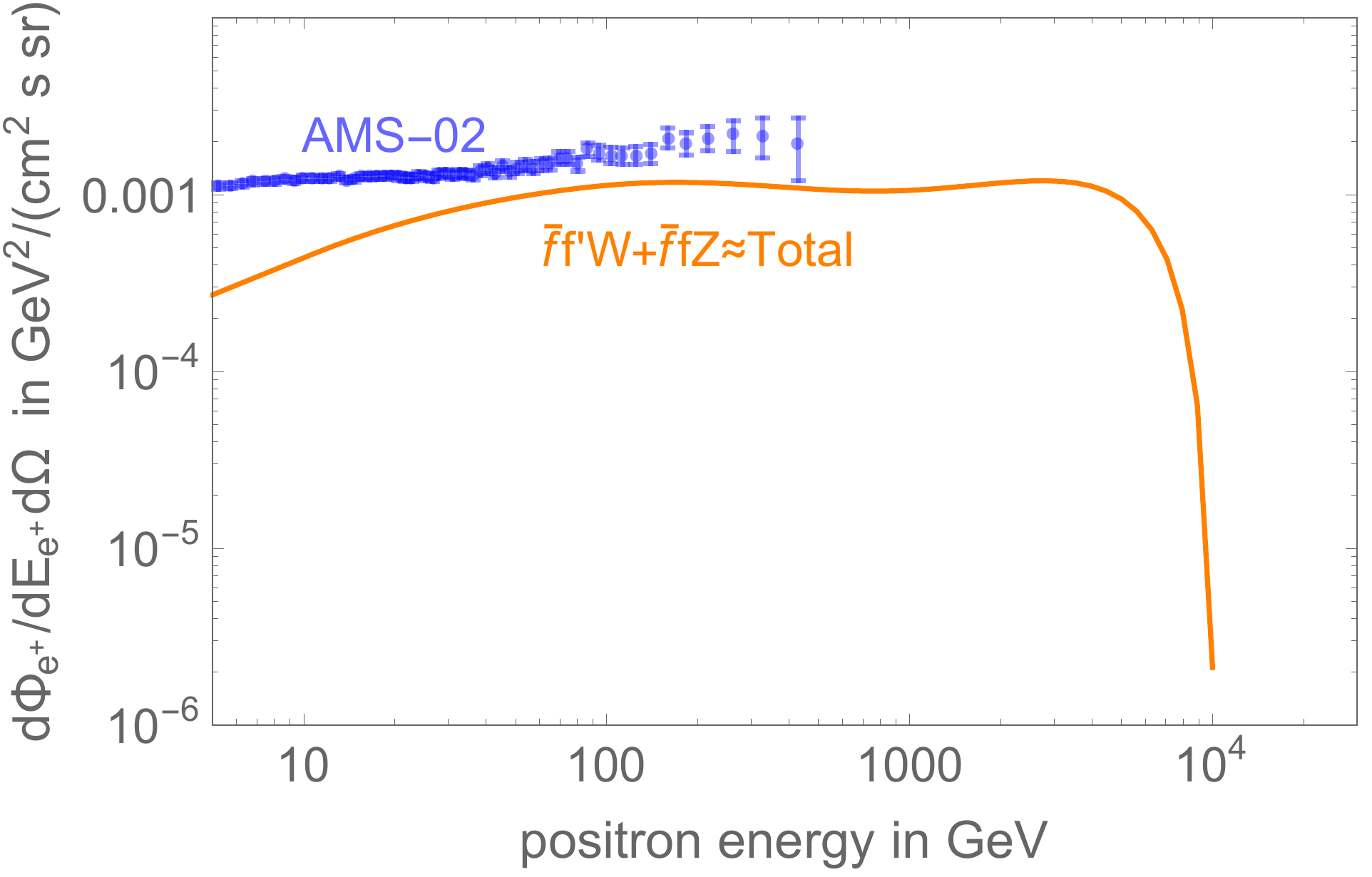}}
\caption{\label{eflux}
Predicted positron flux from decaying DMps contributed by various channels are shown when $\tau=10^{26}~s$. The total flux is shown by the solid line. AMS-02 observations of positron flux are also shown by blue points with error bars~\cite{AMS02}.
}
\end{figure}
\begin{figure}[htbp]
\centering
\includegraphics[scale=0.7]{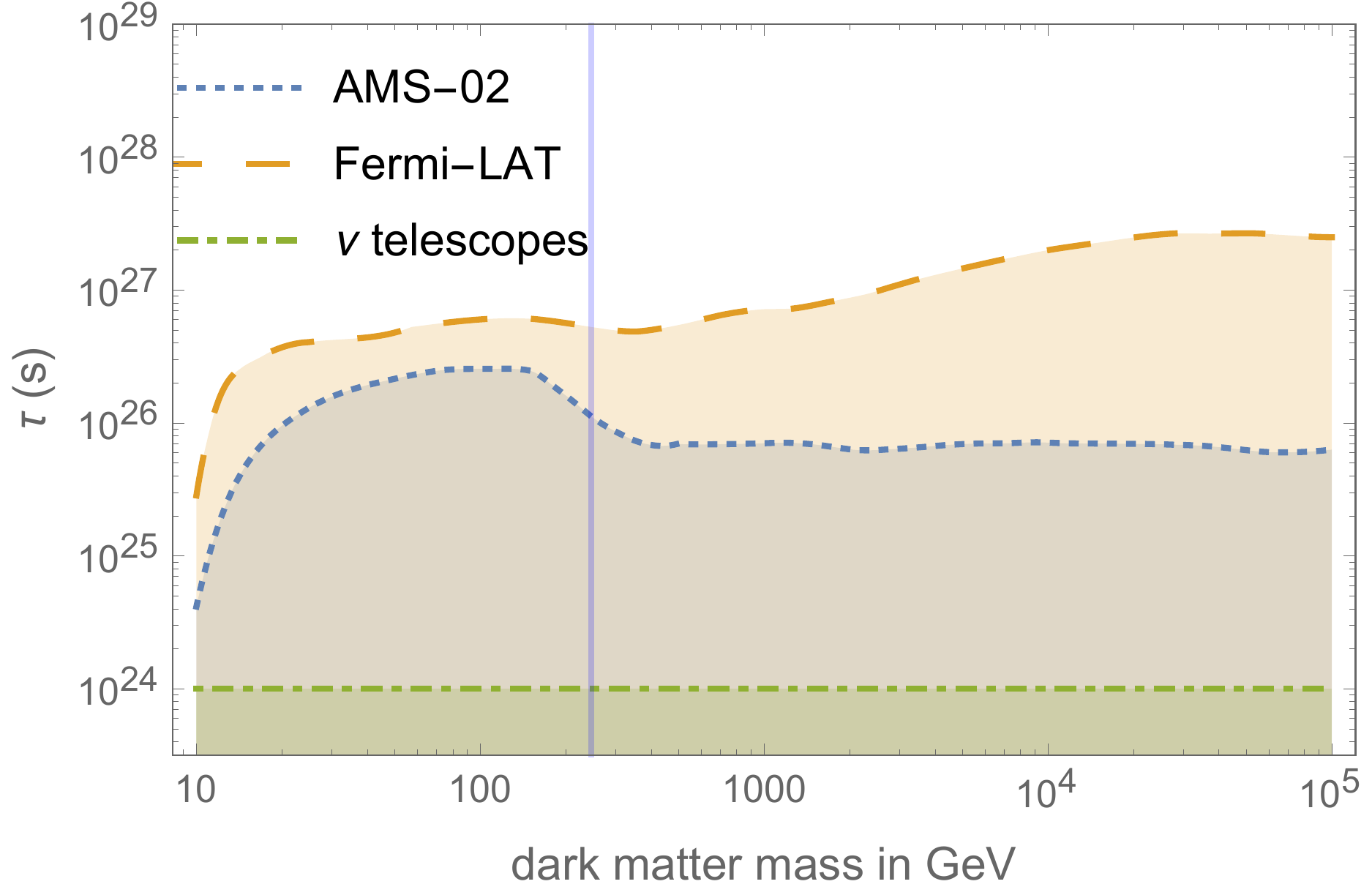}
\caption{\label{mxi2} The $\tau-m_\varphi$ plane. The shadowed regions are regions excluded by observation of the IGRB by Fermi-LAT and the cosmic-ray positron spectrum obtained by AMS-02. For comparison, the conservative excluded parameter space from observations of the cosmic neutrino flux~\cite{Oscar}~\cite{DMdecayTGP} is shown by the shadowed area below the dot-dashed line.}
\end{figure}
\begin{figure}[htbp]
\centering
\includegraphics[scale=0.7]{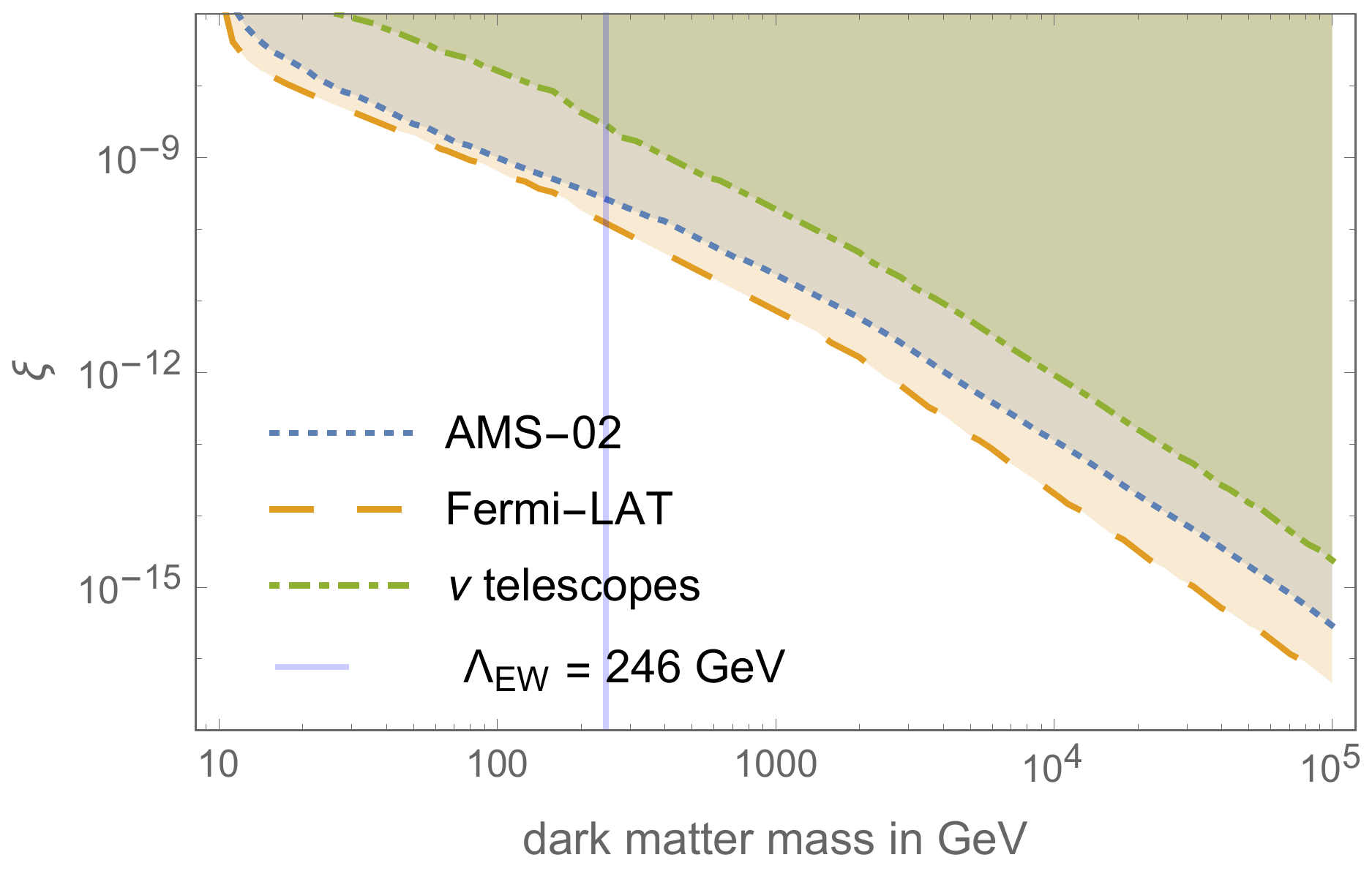}
\caption{\label{mxi1} The $\xi-m_\varphi$ plane. The shadowed regions are regions excluded by observation of the IGRB by Fermi-LAT and the cosmic-ray positron spectrum obtained by AMS-02. For comparison, a conservative excluded parameter space from observations of the cosmic neutrino flux~\cite{Oscar}~\cite{DMdecayTGP} is shown by the shadowed area above the dot-dashed line.}
\end{figure}

The IGRB could be contributed by many unresolved sources, such as non--blazar active galactic nuclei, the unresolved star--forming galaxies, BL Lacertae objects, flat--spectrum radio quasar blazars and electromagnetic cascades generated through ultra--high energy cosmic--ray propagation.
When the IGRB is used to constrain the lifetime of DM, some work takes the contribution of these sources into account, so they could get the most stringent constraints~\cite{CfIGRB}. On the contrary, some work did not take the contribution of these sources into account, so they could get conservative constraints~\cite{chi2test2}.
In our work, we adopt the latter attitude that those contributions are not added in the total flux, so the results we obtained would also be very conservative.

Similar to the IGRB case, it is usually believed that the cosmic positron spectrum has a power--law background. We also do not take into account this contribution in the total predicted flux, so the results obtained by using the cosmic positron flux would also be very conservative.

\section{Results}\label{Results}

Based on the procedure outlined in Section~\ref{decayspectrumGSB}, the photon flux and the positron flux arising from DMp decay, and which would be detected by satellites, were calculated.

Fig.~\ref{1gammaflux} shows the averaged photon flux ($|b|>20^\circ$) from decaying DMps of prompt emission component, extragalactic component, inverse Compton scattering component, and the total flux when the lifetime of ssDM is $\tau=5.3\times10^{26}~s$ and the minUV model is adopted. It can be seen from Fig.~\ref{1gammaflux}(a)-(c) that when $v<m_\varphi<1000~\text{GeV}$, prompt photon flux contributed main part of the total flux, and the contribution of inverse Compton scattering is the least. When the mass of ssDM is large enough (e.g. $m_\varphi=20~\text{TeV}$ as shown in Fig.~\ref{1gammaflux}(d)), the contribution from inverse Compton scattering in the low energy region of the photon spectrum is comparable with the prompt emission component and the extragalactic component.

Fig.~\ref{2gammaflux} show the averaged photon flux ($|b|>20^\circ$) from decaying DMps contributed by different channels when $\tau=5.3\times10^{26}~s$ and the minUV model is adopted. It can be seen from Fig.~\ref{2gammaflux}(a)--(c) that when $v<m_\varphi<1000~\text{GeV}$, two--body decays are of comparable contributions with the three--body channels. This result is consistent with Fig.~\ref{branchratio} that when $v<m_\varphi<1000~\text{GeV}$, the branching ratio of two--body decays are of comparable contributions with the three--body channels. Among all the channels, the $\varphi\to h,h$ channel is more characteristic, its contribution to the photon flux increases slightly near the cut--off. When the mass of the DM particle is $m_\varphi=20~\text{TeV}$, it could be seen from Fig.~\ref{2gammaflux}(d) that most of the photons comes from the $\varphi\to \bar{f}f'W+\bar{f}fZ$ channel. This result is also consistent with Fig.~\ref{branchratio} that when $4\pi v \lesssim m_\varphi \lesssim 10^5~\text{GeV}$, the branching ratio of ssDM is dominated by the same channel.

Fig.~\ref{3gammaflux} shows the absorption of UV photons with the presence of the UV background compared with no UV background regime when $\tau=5.3\times10^{26}~s$. When $v<m_\varphi<1000~\text{GeV}$, comparing the maximum of these discrepancies as shown in Fig.~\ref{3gammaflux}(a)--(c) with the total flux shown in Fig.~\ref{1gammaflux}(a)--(c) or Fig.~\ref{2gammaflux}(a)--(c), we have $(\Delta\Phi/\Phi)_{\text{max}}\sim 10^{-2}$. When the mass of ssDM is $m_\varphi=20~\text{TeV}$, comparing the maximum of the discrepancy as shown in Fig.~\ref{3gammaflux}(d) with the total flux shown in Fig.~\ref{1gammaflux}(d) or Fig.~\ref{2gammaflux}(d), we have $(\Delta\Phi/\Phi)_{\text{max}}\sim 10^{-1}$. These results show that the absorption from the UV background becomes apparent, when the mass of the DM particles is large.

Fig.~\ref{eflux} shows the positron flux from decaying DMps contributed by various channels when $\tau=10^{26}~s$. Fig.~\ref{eflux}(a) shows that when $m_\varphi=v$, three--body decays tends to contribute positrons in the low energy region, while $\varphi\to WW+ZZ$ channel tends to contribute positrons in high energy region.
It could be infer from Fig.~\ref{eflux}(b) and ~\ref{eflux}(c) that when $500~\text{GeV}<m_\varphi<1000~\text{GeV}$, two--body decays are of comparable contributions with the three--body channels. Similar to the photon case, this result is consistent with Fig.~\ref{branchratio} that when $500~\text{GeV}<m_\varphi<1000~\text{GeV}$, the branching ratio of two--body decays are of comparable contributions with the three--body channels. When the mass of the DM particle is $m_\varphi=20~\text{TeV}$, it could be seen from Fig.~\ref{eflux}(d) that most of the positrons comes from the $\varphi\to \bar{f}f'W+\bar{f}fZ$ channel. As expected, this result is consistent with Fig.~\ref{branchratio} that when $4\pi v \lesssim m_\varphi \lesssim 10^5~\text{GeV}$, the branching ratio of ssDM is dominated by the same channel.

Based on the procedure outlined in Section~\ref{ConstraintsfromAMSandFermiLAT}, the excluded two-dimensional parameter space ($\tau,m_\varphi$) is shown in Fig.~\ref{mxi2}, where minUV model is adopted. The shadowed area below the dashed line is the excluded region of parameter space $(\tau, m_\varphi)$ as constrained by Fermi-LAT. The shadowed area below the dotted line is the parameter space $(\tau, m_\varphi)$ excluded by AMS-02. For comparison, a conservative excluded parameter space from observations of the cosmic neutrino flux~\cite{Oscar}~\cite{DMdecayTGP} is shown by the shadowed area above the dot--dashed line. Besides, we also plotted the line for $\Lambda_{\text{EW}}=246$~GeV, which represents the typical energy of the electroweak scale. It could read off that if the mass of the DMp is around the electroweak scale, the lifetime of ssDM smaller than $5.3\times10^{26}~s$ can be excluded. Since $\xi$ reveals the influence of gravity on the global symmetry of ssDM, the excluded region of parameter space $(\xi, m_\varphi)$ is given in Fig.~\ref{mxi1}, where minUV model is adopted.

\section{Discussion and Conclusions}\label{DandC}

Global symmetry can guarantee the stability of ssDM particles. However, the nonminimal coupling between ssDM and gravity can destroy their global symmetry, hence leading to their decay.

In this study, we set constraints on the lifetime and the symmetry breaking strength of ssDM particles using the most sensitive observations of photons and cosmic rays respectively made by Fermi-LAT and AMS-02. The results in Fig.~\ref{mxi1} show that the non--minimal coupling constant between the Ricci scalar and the ssDM receives stronger constraints from indirect detection when the mass of ssDM is larger. This behaviour is attributed to the fact that an ssDM particle with a larger mass has more decay channels and larger phase space. And it confirms O. Cat\`{a} et al.'s conclusion in GeV--TeV range that the exclusion of large regions of the parameter spaces means an additional stabilizing symmetry should be in place.

Different from the previous work by \cite{Sharpfeature}, the mass range of ssDM particles considered in our study is around the GeV--TeV range. Near this scale, the decay channels are more abundant and the phase space is larger. In the work by O. Cat\`{a} et al.~\cite{Sharpfeature}, the lifetime of an ssDM candidate with mass around $m_\varphi\gtrsim~1~\text{MeV}$ decaying through gravity portal is constrained to $\tau\gtrsim~10^{24}-10^{26}~\text{s}$. In this work, the lifetime of ssDM smaller than $5.3\times10^{26}$ is excluded at 3--$\sigma$ confidence level when the mass of the ssDM is around the electroweak scale (246 GeV). The mass region analyzed here contains abundant decay channels that the MeV scale does not have, so the analysis of the decay properties are more comprehensive.

What is going on in parallel with this work is a new paper on this topic\footnote{F. Bezrukov et al., arXiv:2006.03431}, which points out that the fermionic fields should be conformally rescaled in the Einstein Frame. In their regime, all the vertices containing only one gauge boson disappear. Meanwhile, the decay rate of all other channels, including the $\varphi\to\bar{f}_i,f_i$ channels, remains unchanged at tree level. Consequently, in the vicinity of the interested electroweak scale (specifically, $v<m_\varphi<1000~\text{GeV}$), the constraints from IGRB on the global symmetry of ssDM is still in the same order of magnitude compared with our scenario. However, in the region deviating from the electroweak scale (specifically, $m_\varphi<v$ and $10^3~\text{GeV}<m_\varphi<10^5~\text{GeV}$), the constraints from IGRB on the global symmetry of ssDM is significantly weakened. Another parallel related work is also in progress\footnote{H. Azri et al., arXiv:2007.09681}, in which they have developed a different framework for non-minimal couplings of DM through gravity. The framework is, somehow, different from the one developed by O. Cat\`{a} et al. in a sense that the only allowed DM is scalar, and couples only to the Standard Model Higgs boson. Therefore, any decay of DM is through the Higgs (either on-shell or off-shell).

In the context of indirect searches of DM in the galactic region and beyond, many theoretical and experimental works are in progress. On the theoretical side, S. Amoroso et al.~\cite{updateToThePPPC4DM} have produced the spectra within \textsc{Pythia 8.2} (which can be considered as an update to the PPPC 4 DM tables) following several improvements on both the tunings of \textsc{Pythia 8} event generator and the perturbative machinery. Moreover, they have estimated for the first time QCD uncertainties on particle spectra from showering and hadronisation which could be very useful in the context of global fits. On the experimental side, the DAMPE detector was designed to run for at least three years, and the energies measured in the future may be up to about 10~TeV~\cite{DAMPE}. The Large High Altitude Air Shower Observatory (LHAASO) also will be able to detect $\gamma$-ray signals from DM particles of PeV-EeV masses decaying on the time scale up to $3\times10^{29}$ s\footnote{
A. Neronov, D. Semikoz, arXiv:2001.11881}. All these missions can help us further investigate the impact of gravity on DM.

\end{document}